\documentclass[aps,prb,preprint]{revtex4}
\usepackage{amsmath,amssymb}
\usepackage{graphicx}
\usepackage{epsfig}
\usepackage{color}
\usepackage{bm}

\begin{document}
\draft
\title{Interplay of Lorentz-Berry forces in position-momentum spaces for valley-dependent impurity scattering in $\alpha$-$T_3$ lattices}
	
\author{Danhong Huang$^{1}$, Andrii Iurov$^{2}$, Hong-Ya Xu$^{3}$, Ying-Cheng Lai$^{3,4}$ and Godfrey Gumbs$^{5}$}
\affiliation{$^{1}$Air Force Research Laboratory, Space Vehicles Directorate, Kirtland Air Force Base, NM 87117, USA\\
$^{2}$Center for High Technology Materials, University of New Mexico, 1313 Goddard St SE, Albuquerque, NM 87106, USA\\
$^{3}$School of Electrical, Computer and Energy Engineering, Arizona State University, Tempe, AZ 85287, USA\\
$^{4}$Department of Physics, Arizona State University, Tempe, AZ 85287, USA\\
$^{5}$Department of Physics and Astronomy, Hunter College of the City University of New York, 695 Park Avenue, New York, NY 10065, USA}

\date{\today}
	
\begin{abstract}
The Berry-phase mediated valley-selected skew scattering in $\alpha$-$T_3$ lattices is demonstrated.
The interplay of Lorentz and Berry forces in position and momentum spaces is revealed and analyzed.
Many-body screening of the electron-impurity interaction is taken into account to avoid overestimation of back- and skew-scattering of electrons in the system.
Triplet peak from skew interactions at two valleys is found in near-vertical and near-horizontal
forward- and backward-scattering directions for small Berry phases and low magnetic fields.
Magnetic-field dependence in both non-equilibrium and thermal-equilibrium currents is also presented for valley-dependent longitudinal and transverse transports mediated by a Berry phase.
Mathematically, two Boltzmann moment equations are employed for computing scattering-angle distributions of non-equilibrium skew currents by using microscopic inverse energy- and momentum-relaxation times.
Meanwhile, a valley-dependent unbalanced
thermal-equilibrium anomalous Hall current induced by the Berry force in momentum space, due to different mobilities for two valleys,
is also computed for comparisons.
\end{abstract}
\maketitle

\section{Introduction}
\label{sec-1}

In electronics or spintronics\,\cite{spin}, information is encoded through either charge or spin. Valley quantum numbers, on the other hand, become another way to distinguish and designate quantum states of a crystal lattice,
which leads to the so-called valleytronics\,\cite{ref1,ref2} and has already attracted a lot of interest\,\cite{ref3,ref4,ref5,ref6,ref7,ref8,ref9} from both fundamental research and application perspectives.
Physically speaking, valleytronics bases itself on controlling the valley degree-of-freedom of certain semiconductors with multiple valleys inside their first Brillouin zone, such as
$\mbox{\boldmath$\it{\Gamma}$}$, $\mbox{\boldmath$K$}$, $\mbox{\boldmath$L$}$ and $\mbox{\boldmath$M$}$ band-extreme points.
As a comparison, electron spins have already been used for storing, manipulating and reading out bits of information.\,\cite{spin2}
Therefore, we expect valleytronics will also demonstrate similar functionalities through multiple band extrema, where the information of $0\,$s and $1\,$s could be stored as discrete crystal momenta.
\medskip

By taking graphene\,\cite{graph} as an example,
its two nonequivalent valleys can be described as an ideal two-state system (similar to the isospin degree of freedom), and its two nonequivalent Dirac points, \mbox{\boldmath$K$} and \mbox{\boldmath$K$}$'$
in the first Brillouin zone, are associated with distinct momenta or valley quantum numbers. These two valleys are well separated by a vary large crystal momentum, and therefore become
robust against usual external perturbations at room temperature.
Quantum manipulation of valleys in semiconductors has just been demonstrated recently,\,\cite{valley} and electrons belonging to different valleys are employed for quantum-information processing.
Beyond graphene, valley characteristics are also present in other two-dimensional materials such as silicene, germanene, MoS$_2$, WSe$_2$, and etc.
\medskip

By looking from a technical perspective, a key issue in valleytronics turns out to be the separation of electrons with different valley quantum numbers in either position or momentum space, i.e., the so-called valley filters\,\cite{filter}.
One way to obtain valley filtering is based on the valley Hall effect\,\cite{valley} (VHE), where electrons from different valleys can be separated spatially.
There are other physical phenomena, e.g., the anomalous Hall effect\,\cite{ahe} (AHE) and the spin Hall effect\,\cite{she} (SHE), which are closely related to VHE.
In fact, SHE has already been proven as a connection between the electrical and spin currents and can be used for spin-current generation and detection electrically in spintronics.
In a similar way, we expect VHE can also generate transverse valley currents in position space like SHE.
\medskip

The $\alpha$-$T_3$ physics model is recognized as the most recent and promising candidate for novel two-dimensional materials.
Its low-energy dispersions. including a flat band, can be found from a spin-$1$ particle's Dirac-Weyl Hamiltonian\,\cite{dice,re2} and acquires 
a close similarity when compared with graphene\,\cite{re3,re4,re5}
The experimental observation for a dispersion-less state was confirmed\,\cite{re6,re7} in a photonic Lieb lattice
formed by a two-dimensional array of optical waveguides. This photonic Lieb lattice can support three energy bands, including a perfectly 
flat middle band (i.e., an infinite effective mass). Moreover, these flat-band states are remarkable robustness, 
even in the presence of disorders. Alternatively, the realization of the Lieb lattice can be fulfilled with an optical lattice,\,\cite{re8} 
which has a flat energy band as the first excited state. 
Furthermore, by employing accidental degeneracy, dielectric photonic crystals with zero-refractive-index can be designed and fabricated 
that exhibit Dirac cone dispersion at the center of the Brillouin zone at a finite frequency.\,\cite{re9,re10}
\medskip

\begin{figure}
\centering
\includegraphics[width=0.85\textwidth]{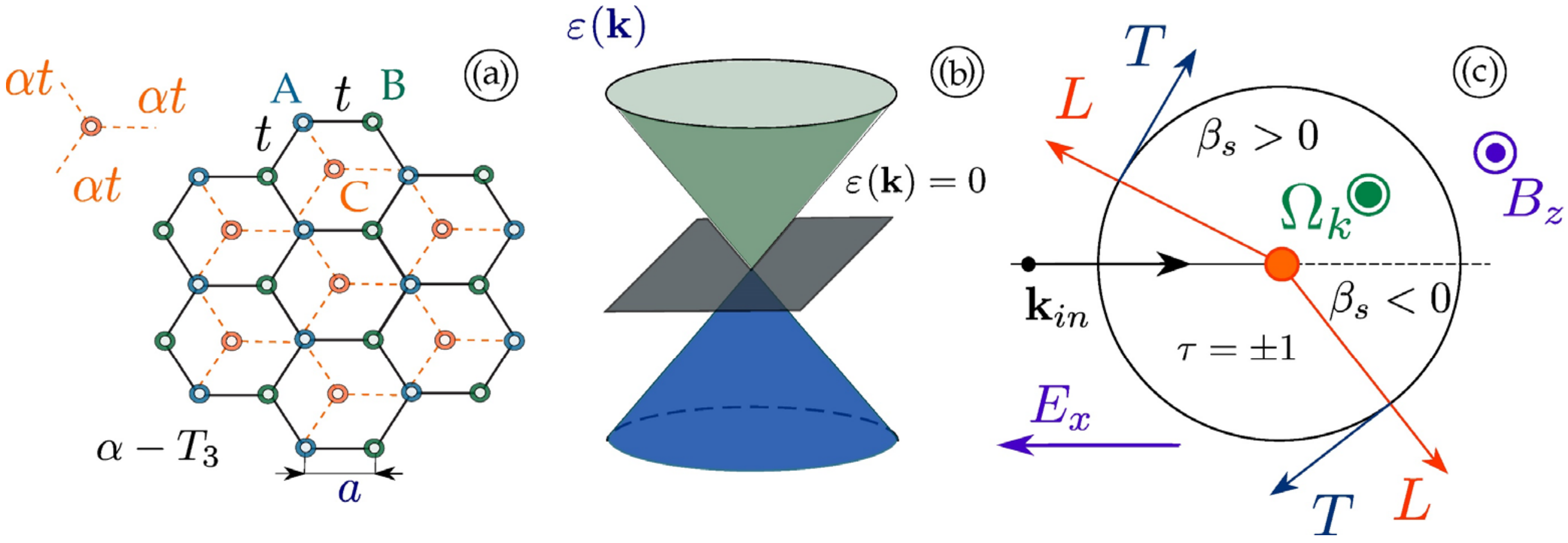}
\caption{(a) $\alpha$-$T_3$ lattice with three atoms ($A$, $B$, $C$) per unit cell within the $(x,y)$-plane, where the $\alpha$ parameter characterizes the ratio of the bonding strengths between $A$-$C$ and $A$-$B$ atoms;
(b) illustration for a band structure featuring three bands of $\alpha$-$T_3$ lattice, where the middle one is flat;
(c)	schematic diagram for a scattering angle $\beta_s$ of an incident electron with wavevector $\mbox{\boldmath$k$}_{\rm in}$ by different
impurities at two valleys characterized by $\tau=\pm 1$ under an applied electric field $\mbox{\boldmath$E$}_x$ along the $x$ direction,
where an external non-quantizing magnetic field $\mbox{\boldmath$B$}_z$, and the internal Berry curvature $\mbox{\boldmath$\Omega$}_k$ as well, are along the $z$ direction and the longitudinal (transverse) scattering is labeled by $L$ ($T$), respectively.}
\label{fig1}
\end{figure}

The idea of highly-efficient valley filtering in $\alpha$-$T_3$ lattices with variable Berry phase, as shown schematically in
Figs.\,\ref{fig1}($a$) and \ref{fig1}($b$), has been reported very recently\,\cite{ycl} with a Berry-phase-mediated VHE,
which is termed as gVHE due to the geometric nature of the underlying mechanism. In this case, the Berry phase in momentum space can be fractionally quantized, and
charge-neutral valley currents occur through skew scattering by the usual thermally-ionized donor or acceptor
impurities. Furthermore, a physical understanding is sought for resonant valley filtering\,\cite{resonant} assisted by skew scattering to ensure gVHE
could be robust against both thermal fluctuations and structural disorders as a result of large inter-valley momentum separation.
\medskip

Since novel two-dimensional (2D) materials span the full range of electronic properties, including insulators, semiconductors, semimetals and metals, we hope to stack them layer by layer through van der waals forces so as to build various compact
planar electronic devices with high and multifunctional performance, light weight, low-power consumption, flexibility, and even transparency.
The semiconducting 2D monolayer gives rise to excellent gate control in field-effect transistors (FETs) with much shorter gate lengths (or smaller and faster transistors).
Furthermore, by aligning the material's low-effective-mass lattice direction with the FET's transport, the carrier mobility will be enhanced greatly along with a high carrier density.
Recent theoretical and experimental endeavors on the charge transfer across a 2D material interface lead to the successful fabrication of low-resistance contacts, where the covalently bonded in-plane interfaces between different 2D materials demonstrate hope for reducing contact resistances, power consumption and heat generation.
\medskip

In this paper, our previous single-particle quantum-mechanical theory\,\cite{ycl} for $\alpha$-$T_3$ lattices with variable Berry phases will be generalized into a many-body quantum-statistical theory based on a generalized Boltzmann transport formalism,
which microscopically calculates the inverse energy-relaxation time using the screened second-order Born approximation, the inverse momentum-relaxation-time tensor for electron elastic scattering by ionized donor and acceptor impurities, and the generalized mobility tensor
based on the force-balance equation. Moreover, the zeroth- and first-order moment equations derived from the general Boltzmann transport equation will be employed for computing both the forward- and backward-scattering (near-horizontal) and skew-scattering (near-vertical) currents.
Furthermore, the interplay between Lorentz and Berry forces acting on electrons in position and momentum space
for both non-equilibrium and thermal-equilibrium currents is analyzed and explained.
\medskip

The rest of this paper is organized as follows. In Sec.\,\ref{sec-2}, we derive the zeroth- and first-order Boltzmann moment equations for calculating both non-equilibrium back- and skew-scattering currents in $\alpha$-$T_3$ lattices
as well as thermal-equilibrium anomalous Hall current. Meanwhile, both energy- and momentum-relaxation times are computed microscopically.
In Sec.\,\ref{sec-3}, we present numerical results for valley-dependent distributions of longitudinal and transverse currents with respect to different scattering directions, and
valley-dependent 2D contour plots for partial back- and skew-scattering currents as a function of both magnetic field and Berry phase at several scattering angles.
We also display in this section the total back- and skew-scattering currents in individual valleys as a function of magnetic field for different Berry phases.
Finally, a summary and some remarks are presented in Sec.\,\ref{sec-4}.

\section{Model and Theory}
\label{sec-2}

For an $n$-doped two-dimensional (2D) $\alpha$-$T_3$ lattice,
we start with the semi-classical Boltzmann transport equation for doped electrons in a conduction band $\varepsilon(k_\|)=\hbar v_Fk_\|$ of this 2D material,
where $v_F$ and $k_\|$ are the Fermi velocity and wave number of electrons.
In this case, the electron distribution function $f_\tau(\mbox{\boldmath$r$}_\|,\mbox{\boldmath$k$}_\|;t)$ in position-momentum spaces satisfies\,\cite{ziman}

\[
\frac{\partial f_\tau(\mbox{\boldmath$r$}_\|,\mbox{\boldmath$k$}_\|;t)}{\partial t}+\Big\langle\frac{d\mbox{\boldmath$r$}_\|(t)}{dt}\Big\rangle_{\rm av}\cdot\mbox{\boldmath$\nabla$}_{{\bf r}_\|}f_\tau(\mbox{\boldmath$r$}_\|,\mbox{\boldmath$k$}_\|;t)
\]
\begin{equation}
+\Big\langle\frac{d\mbox{\boldmath$k$}_\|(t)}{dt}\Big\rangle_{\rm av}\cdot\mbox{\boldmath$\nabla$}_{{\bf k}_\|}f_\tau(\mbox{\boldmath$r$}_\|,\mbox{\boldmath$k$}_\|;t)
=\left.\frac{\partial f_\tau(\mbox{\boldmath$r$}_\|,\mbox{\boldmath$k$}_\|;t)}{\partial t}\right|_{\rm coll}\ ,\ \
\label{eqn-1}
\end{equation}
where $\tau=\pm 1$ characterize two inequivalent valleys $\mbox{\boldmath$K$}$ and $\mbox{\boldmath$K$}'$,
$\mbox{\boldmath$r$}_\|=\{x,y\}$ and $\mbox{\boldmath$k$}_\|=\{k_x,k_y\}$ are 2D position and wave vector, respectively.
The term on the right-hand side of Eq.\,(\ref{eqn-1}) corresponds to all collision contributions of electrons with ionized-impurities, phonons, other electrons, etc. Moreover, for electrons,
we get their group velocities through $\mbox{\boldmath$v$}(\mbox{\boldmath$k$}_\|)=(1/\hbar)\mbox{\boldmath$\nabla$}_{{\bf k}_\|}\varepsilon(k_\|)=(\mbox{\boldmath$k$}_\|/k_\|)\,v_F$.
Meanwhile, we find semi-classically that\,\cite{niu}
$\langle d\mbox{\boldmath$r$}_\|(t)/dt\rangle_{\rm av}=\mbox{\boldmath$v$}(\mbox{\boldmath$k$}_\|)-d\bar{\mbox{\boldmath$K$}}_0(t)/dt\times\mbox{\boldmath$\Omega$}_\perp(\mbox{\boldmath$k$}_\|)\equiv\mbox{\boldmath$v$}^\ast(\mbox{\boldmath$k$}_\|,t)$, where
$\mbox{\boldmath$v$}^\ast(\mbox{\boldmath$k$}_\|,t)$ contains the so-called anomalous group velocity\,\cite{niu-book},
$\bar{\mbox{\boldmath$K$}}_0(t)$ is the center-of-mass wave vector,
$\mbox{\boldmath$\Omega$}_\perp(\mbox{\boldmath$k$}_\|)=\mbox{\boldmath$\nabla$}_{{\bf k}_\|}\times\bar{\mbox{\boldmath$R$}}_0(\mbox{\boldmath$k$}_\|)$ is called the Berry curvature,
and $\bar{\mbox{\boldmath$R$}}_0(\mbox{\boldmath$k$}_\|)=\langle\mbox{\boldmath$k$}_\||\hat{\mbox{\boldmath$r$}}_\||\mbox{\boldmath$k$}_\|\rangle=\langle \mbox{\boldmath$k$}_\||i\hat{\mbox{\boldmath$\nabla$}}_{{\bf k}_\|}|\mbox{\boldmath$k$}_\|\rangle$ is called the Berry connection and related to the quantum-mechanical average of the center-of-mass position operator
with respect to Bloch states $|\mbox{\boldmath$k$}_\|\rangle$ of a conduction band under the adiabatic condition\,\cite{niu-book}.
Furthermore, we introduce a semi-classical Newton-type force equation\,\cite{ziman} for the wave vector of electrons, yielding
$\langle d\mbox{\boldmath$k$}_\|(t)/dt\rangle_{\rm av}=(1/\hbar)\langle\mbox{\boldmath$F$}_{\rm em}(\mbox{\boldmath$k$}_\|,t)\rangle_{\rm av}
=-(e/\hbar)\Big\langle\left[\mbox{\boldmath$E$}_\|(t)+\mbox{\boldmath$v$}(\mbox{\boldmath$k$}_\|)\times\mbox{\boldmath$B$}_\perp(t)\right]\Big\rangle_{\rm av}$,
where $\mbox{\boldmath$E$}_\|(t)$ and $\mbox{\boldmath$B$}_\perp(t)$ are external time-dependent electric and magnetic fields, respectively, and $\mbox{\boldmath$F$}_{\rm em}(\mbox{\boldmath$k$}_\|,t)$ is the electromagnetic force acting on an electron in the $\mbox{\boldmath$k$}_\|$ state. Here, $\mbox{\boldmath$B$}_\perp(t)$ is assumed as a non-quantizing magnetic field with Landau-level separation $\sim\hbar\omega_c$ smaller than the level lifetime broadening $\hbar/\bar{\tau}$.
\medskip

Based on Eq.\,(\ref{eqn-1}), the zeroth-order Boltzmann moment equation\,\cite{jmo,backes} can be obtained simply by summing over all $\mbox{\boldmath$k$}_\|$ states on both sides of this equation.
After ignoring the inter-valley scattering at low temperatures with a very large transition momentum, this gives rise to the electron number conservation equation, i.e.,
$\partial\rho/\partial t+\mbox{\boldmath$\nabla$}_{{\bf r}_\|}\cdot\mbox{\boldmath$J$}=0$,
where the number of electrons $\rho(\mbox{\boldmath$r$}_\|,t)$ per area, as well as the particle-number current $\mbox{\boldmath$J$}(\mbox{\boldmath$r$}_\|,t)$ per length, are defined by
$\displaystyle{\rho(\mbox{\boldmath$r$}_\|,t)=\frac{2}{{\cal S}}\sum\limits_{\tau,{\bf k}_\|}\,f_\tau(\mbox{\boldmath$r$}_\|,\mbox{\boldmath$k$}_\|;t)}$ and
$\displaystyle{\mbox{\boldmath$J$}(\mbox{\boldmath$r$}_\|,t)=\frac{2}{{\cal S}}\sum\limits_{\tau,{\bf k}_\|}\,\mbox{\boldmath$v$}^\ast(\mbox{\boldmath$k$}_\|,t)\,f_\tau(\mbox{\boldmath$r$}_\|,\mbox{\boldmath$k$}_\|;t)}$
with ${\cal S}$ as the sheet area.
\medskip

For the first-order Boltzmann moment equation, on the other hand,
we have to employ the so-called Fermi kinetics\,\cite{jmo,backes}. For this purpose, we
first introduce the energy-relaxation-time approximation for collisions, given explicitly by

\begin{equation}
\left.\frac{\partial f_\tau(\mbox{\boldmath$r$}_\|,\mbox{\boldmath$k$}_\|;t)}{\partial t}\right|_{\rm coll}=-\,\frac{f_\tau(\mbox{\boldmath$r$}_\|,\mbox{\boldmath$k$}_\|;t)-f_T^{(0)}[\varepsilon(k_\|)]}{\tau_\phi(\mbox{\boldmath$k$}_\|,\tau)}\ ,
\label{eqn-2}
\end{equation}
which conserves the particle number, 
where $f_T^{(0)}(x)=\{1+\exp[(x-u_0)/k_BT]\}^{-1}$ is the Fermi function for electrons in thermal-equilibrium states,
$T$ is the sample temperature, $u_0(T)$ is the chemical potential for doped electrons, and $\tau_\phi(\mbox{\boldmath$k$}_\|,\tau)$ is the microscopic and valley-dependent energy-relaxation time for electrons in the $\mbox{\boldmath$k$}_\|$ state.
The detailed quantum-statistical calculation of $\tau_\phi(\mbox{\boldmath$k$}_\|,\tau)$ can be found in Appendix\ \ref{app-1}.
The chemical potential $u_0(T)$ of a canonical system should be determined self-consistently by the constraint:
$\displaystyle{4\sum\limits_{{\bf k}_\|}\,f_T^{(0)}[\varepsilon(k_\|)]=\int d^2\mbox{\boldmath$r$}_\|\,\rho(\mbox{\boldmath$r$}_\|,t)\equiv\frac{2}{{\cal S}}\sum\limits_{\tau,{\bf k}_\|}
\int d^2\mbox{\boldmath$r$}_\|\,f_\tau(\mbox{\boldmath$r$}_\|,\mbox{\boldmath$k$}_\|;t)=N_0=\rho_0{\cal S}}$,
where $N_0$ and $\rho_0$ represent the fixed total number of spin-degenerate electrons and the electron areal density.
Finally, applying this energy relaxation-time approximation to Eq.\,(\ref{eqn-1}), we arrive at

\[
f_\tau(\mbox{\boldmath$r$}_\|,\mbox{\boldmath$k$}_\|;t)+\bar{\tau}_\phi(T,\tau)\,\frac{\partial f_\tau(\mbox{\boldmath$r$}_\|,\mbox{\boldmath$k$}_\|;t)}{\partial t}\approx f_T^{(0)}[\varepsilon(k_\|)]\,
-\frac{\bar{\tau}_\phi(T,\tau)}{\hbar}\,
\langle\mbox{\boldmath$F$}_{\rm em}(\mbox{\boldmath$k$}_\|,t)\rangle\cdot\mbox{\boldmath$\nabla$}_{{\bf k}_\|}f_T^{(0)}[\varepsilon(k_\|)]
\]
\begin{equation}
-\bar{\tau}_\phi(T,\tau)\,\mbox{\boldmath$v$}^\ast({\bf k}_\|)\cdot\mbox{\boldmath$\nabla$}_{{\bf r}_\|}f_T^{(0)}[\varepsilon(k_\|)]
=f_T^{(0)}[\varepsilon(\mbox{\boldmath$k$}_\|)]
-\frac{\bar{\tau}_\phi(T,\tau)}{\hbar}\langle\mbox{\boldmath$F$}(\mbox{\boldmath$k$}_\|,t)\rangle\cdot\mbox{\boldmath$\nabla$}_{{\bf k}_\|}f_T^{(0)}[\varepsilon(\mbox{\boldmath$k$}_\|)]\ ,
\label{eqn-3}
\end{equation}
where we have assumed $T$ and $u_0$ are spatially-uniform within the sample, and the thermally-averaged
and valley-dependent energy-relaxation time $\bar{\tau}_\phi(T,\tau)$ is defined by
$\displaystyle{\frac{1}{\bar{\tau}_\phi(T,\tau)}=\frac{2}{N_0}\sum\limits_{{\bf k}_\|}\,\frac{f_T^{(0)}[\varepsilon(k_\|)]}{\tau_\phi(\mbox{\boldmath$k$}_\|,\tau)}}$.
By introducing another microscopic inverse momentum-relaxation-time tensor $\tensor{\mbox{\boldmath$\cal T$}}_p^{-1}(\tau,\phi)$, we can further rewrite the force-balance equation\,\cite{huang} for the macroscopic center-of-mass wave vector $\mbox{\boldmath$K$}^{\tau,\phi}_0(t)$ in steady states as

\[
\frac{d\mbox{\boldmath$K$}^{\tau,\phi}_0(t)}{dt}=-\tensor{\mbox{\boldmath$\cal T$}}_p^{-1}(\tau,\phi)\cdot\mbox{\boldmath$K$}^{\tau,\phi}_0(t)+\frac{1}{\hbar}\,\mbox{\boldmath$F$}_{\tau,\phi}(t)
\]
\begin{equation}
= -\tensor{\mbox{\boldmath$\cal T$}}_p^{-1}(\tau,\phi)\cdot\mbox{\boldmath$K$}^{\tau,\phi}_0(t)
-\frac{e}{\hbar}\,\left\{\mbox{\boldmath$E$}_\|(t)+\left(\frac{v_F}{k_F}\right)\mbox{\boldmath$K$}^{\tau,\phi}_0(t)\times\mbox{\boldmath$B$}_\perp(t)\right\}=0\ ,
\label{eqn-4}
\end{equation}
where $\mbox{\boldmath$F$}_{\tau,\phi}(t)\equiv\langle\mbox{\boldmath$F$}_{\rm em}(\mbox{\boldmath$k$}_\|,t)\rangle_{\rm av}
= -e\left\{\mbox{\boldmath$E$}_\|(t)+\left(v_F/k_F\right)\mbox{\boldmath$K$}^{\tau,\phi}_0(t)\times\mbox{\boldmath$B$}_\perp(t)\right\}$ is the macroscopic electromagnetic force,
and $k_F=\sqrt{\pi\rho_0}$ is the Fermi wave number. The detailed quantum-statistical calculation of the inverse momentum-relaxation-time tensor $\tensor{\mbox{\boldmath$\cal T$}}_p^{-1}(\tau,\phi)$ is provided in Appendix\ \ref{app-2}.
The solution of Eq.\,(\ref{eqn-4}) can be formally expressed as $\mbox{\boldmath$K$}^{\tau,\phi}_0(t)=\left(k_F/v_F\right)\,\tensor{\mbox{\boldmath$\mu$}}_{\tau,\phi}(\mbox{\boldmath$B$}_\perp(t),\,\tensor{\mbox{\boldmath$\cal T$}}_p^{-1})\cdot\mbox{\boldmath$E$}_\|(t)$,
where $\tensor{\mbox{\boldmath$\mu$}}_{\tau,\phi}(\mbox{\boldmath$B$}_\perp,\tensor{\mbox{\boldmath$\cal T$}}_p^{-1})$ is the so-called mobility tensor of electrons.
The details for calculating the steady-state mobility tensor $\tensor{\mbox{\boldmath$\mu$}}_{\tau,\phi}(\mbox{\boldmath$B$}_\perp,\tensor{\mbox{\boldmath$\cal T$}}_p^{-1})$ are presented in Appendix \ref{app-3}.
Using this mobility tensor, we can simply write $\mbox{\boldmath$F$}_{\tau,\phi}(t)=\left(\hbar k_F/v_F\right)\tensor{\mbox{\boldmath$\cal T$}}_p^{-1}(\tau,\phi)\cdot\left\{\tensor{\mbox{\boldmath$\mu$}}_{\tau,\phi}(\mbox{\boldmath$B$}_\perp(t),\,\tensor{\mbox{\boldmath$\cal T$}}_p^{-1})\cdot\mbox{\boldmath$E$}_\|(t)\right\}$.
\medskip

In a similar way in deriving the zeroth-order Boltzmann moment equation,
multiplying both sides of Eq.\,(\ref{eqn-3}) by $\mbox{\boldmath$v$}^\ast(\mbox{\boldmath$k$}_\|,t)$ and summing over all electron $\mbox{\boldmath$k$}_\|$ states afterwards, we are left with the following dynamical equation

\[
\mbox{\boldmath$J$}_{\tau,\phi}(t)+\bar{\tau}_\phi(T,\tau)\,\frac{\partial \mbox{\boldmath$J$}_{\tau,\phi}(t)}{\partial t}
=\frac{2}{{\cal S}}\,\sum_{{\bf k}_\|}\,\mbox{\boldmath$v$}^\ast(\mbox{\boldmath$k$}_\|,t)\,f_T^{(0)}[\varepsilon(k_\|)]
\]
\[
-\bar{\tau}_\phi(T,\tau)\,
\frac{2}{{\cal S}}\,\sum_{{\bf k}_\|}\,\mbox{\boldmath$v$}^\ast(\mbox{\boldmath$k$}_\|,t)
\left[\mbox{\boldmath$F$}_{\tau,\phi}(t)\cdot\mbox{\boldmath$v$}(\mbox{\boldmath$k$}_\|)\right]\,\frac{\partial f_T^{(0)}[\varepsilon(k_\|)]}{\partial\varepsilon}
\]
\[
=\frac{2e}{\hbar{\cal S}}\,\sum_{{\bf k}_\|}\,\left\{\left[\mbox{\boldmath$E$}_\|(t)+\left(\tensor{\mbox{\boldmath$\mu$}}_{\tau,\phi}(\mbox{\boldmath$B$}_\perp(t),\,\tensor{\mbox{\boldmath$\cal T$}}_p^{-1})\cdot\mbox{\boldmath$E$}_\|(t)\right)\times\mbox{\boldmath$B$}_\perp(t)\right]\times\mbox{\boldmath$\Omega$}_\perp(\mbox{\boldmath$k$}_\|)\right\}\,f_T^{(0)}[\varepsilon(k_\|)]
\]
\[
+\bar{\tau}_\phi(T,\tau)\,\left(\frac{\hbar k_F}{v_F}\right)\,\frac{2}{{\cal S}}\,\sum_{{\bf k}_\|}\,\mbox{\boldmath$v$}(\mbox{\boldmath$k$}_\|)\,
\]
\begin{equation}
\times\left\{\tensor{\mbox{\boldmath$\cal T$}}_p^{-1}(\tau,\phi)\cdot\left[\tensor{\mbox{\boldmath$\mu$}}_{\tau,\phi}(\mbox{\boldmath$B$}_\perp(t),\,\tensor{\mbox{\boldmath$\cal T$}}_p^{-1})\cdot\mbox{\boldmath$E$}_\|(t)\right]\right\}\cdot\mbox{\boldmath$v$}(\mbox{\boldmath$k$}_\|)\left\{-\frac{\partial f_T^{(0)}[\varepsilon(k_\|)]}{\partial\varepsilon}\right\}\ ,
\label{eqn-5}
\end{equation}
where the second term on the left-hand side of the equation results from the non-adiabatic correction to the macroscopic particle-number current $\mbox{\boldmath$J$}_{\tau,\phi}(t)$ per length.
From Eq.\,(\ref{eqn-5}) we know $\mbox{\boldmath$J$}_{\tau,\phi}$
is also independent of $\mbox{\boldmath$r$}_\|$ within our approximation. As a result, from the electron number conservation equation,
we find the number of electrons $\rho$ per area
must be a constant $\rho_0$, determined by $\displaystyle{\rho_0=\frac{4}{{\cal S}}\sum\limits_{{\bf k}_\|}\,f_T^{(0)}[\varepsilon(k_\|)]}$,
which determines the chemical potential $u_0(T)$ of the sample at any given temperature $T$.
\medskip

If $T$ is low, i.e., $\displaystyle{-\frac{\partial f_T^{(0)}[\varepsilon(k_\|)]}{\partial\varepsilon}\approx\delta\left(E_F-\varepsilon(k_\|)\right)}$, and external fields are assumed static
$\mbox{\boldmath$E$}^\|_0$ and $\mbox{\boldmath$B$}^\perp_0$, we get from Eq.\,(\ref{eqn-5})
the total charge ($-e$) current $\mbox{\boldmath$j$}(\tau,\phi)=\mbox{\boldmath$j$}_{1}(\tau,\phi)+\mbox{\boldmath$j$}_2(\tau,\phi)$ per length for each valley, where
$E_F=\hbar v_Fk_F$ is the Fermi energy of electrons. Explicitly, we calculate the two current components $\mbox{\boldmath$j$}_{1}(\tau,\phi)$ and $\mbox{\boldmath$j$}_2(\tau,\phi)$ as

\[
\mbox{\boldmath$j$}_1(\tau,\phi)=-\frac{ek_F^2\bar{\tau}_\phi(k_F,\tau)}{2\pi^2v_F^2}\int\limits_0^{2\pi} d\theta_{{\bf k}_\|}\,\mbox{\boldmath$v$}(\theta_{{\bf k}_\|})
\left\{\mbox{\boldmath$\cal T$}_p^{-1}(k_F,\tau,\phi)\cdot\left[\tensor{\mbox{\boldmath$\mu$}}_{\tau,\phi}(\mbox{\boldmath$B$}^\perp_0,\tensor{\mbox{\boldmath$\cal T$}}_p^{-1})
\cdot\mbox{\boldmath$E$}^\|_0\right]\right\}\cdot\mbox{\boldmath$v$}(\theta_{{\bf k}_\|})
\]
\begin{equation}
=-\frac{ek_F^2\bar{\tau}_\phi(k_F,\tau)}{2\pi}\int\limits_{-\pi}^\pi d\beta_s\left[\hat{\mbox{\boldmath$e$}}_x{\cal C}_x(k_F,\tau,\phi,\beta_s)+\hat{\mbox{\boldmath$e$}}_y{\cal C}_y(k_F,\tau,\phi,\beta_s)\right]
\equiv\int\limits_{-\pi}^\pi d\beta_s\,\tilde{\mbox{\boldmath$j$}}_1(\tau,\phi,\beta_s)\ ,
\label{eqn-6}
\end{equation}
which is mediated by the Lorentz force in position space, and

\[
\mbox{\boldmath$j$}_2(\tau,\phi)=-\frac{e^2}{2\pi^2\hbar}\int d^2\mbox{\boldmath$k$}_\|\,\Theta(k_F-k_\|)\,
\left\{\left[\mbox{\boldmath$E$}_0^\|
+\left(\tensor{\mbox{\boldmath$\mu$}}_{\tau,\phi}(\mbox{\boldmath$B$}_0^\perp,\tensor{\mbox{\boldmath$\cal T$}}_p^{-1})\cdot\mbox{\boldmath$E$}_0^\|\right)\times\mbox{\boldmath$B$}_0^\perp\right]
\times\mbox{\boldmath$\Omega$}_\perp(\mbox{\boldmath$k$}_\|)\right\}
\]
\[
=-\frac{e^2}{2\pi^2\hbar}\int d^2\mbox{\boldmath$k$}_\|\,\Theta(k_F-k_\|)\,
\left\{\hat{\mbox{\boldmath$e$}}_x\left[E_y-B_z\left(\mu_{xx}(k_F,\tau,\phi)\,E_x+\mu_{xy}(k_F,\tau,\phi)\,E_y\right)\right]\,\Omega_{\tau,\phi}(\mbox{\boldmath$k$}_\|)\right.
\]
\[
\left.-\hat{\mbox{\boldmath$e$}}_y\left[E_x+B_z\left(\mu_{yx}(k_F,\tau,\phi)\,E_x+\mu_{yy}(k_F,\tau,\phi)\,E_y\right)\right]\,\Omega_{\tau,\phi}(\mbox{\boldmath$k$}_\|)\right\}
\]
\[
=-\frac{e^2}{2\pi^2\hbar}\left\{\frac{\tau(1-\alpha^2)\pi}{1+\alpha^2}\right\}
\left\{\hat{\mbox{\boldmath$e$}}_x\left[E_y-B_z\left(\mu_{xx}(k_F,\tau,\phi)\,E_x+\mu_{xy}(k_F,\tau,\phi)\,E_y\right)\right]\right.
\]
\begin{equation}
\left.-\hat{\mbox{\boldmath$e$}}_y\left[E_x+B_z\left(\mu_{yx}(k_F,\tau,\phi)\,E_x+\mu_{yy}(k_F,\tau,\phi)\,E_y\right)\right]\right\}
\equiv j_{2x}(\tau,\phi)\,\hat{\mbox{\boldmath$e$}}_x+j_{2y}(\tau,\phi)\,\hat{\mbox{\boldmath$e$}}_y\ ,
\label{eqn-6.2}
\end{equation}
which is mediated by the Berry curvature (or Berry force) in momentum space.
Here, $\Theta(x)$ is a unit-step function, $\mu_{ij}(k_F,\tau,\phi)$ for $i,j=x,y$ are four elements of the mobility tensor $\tensor{\mbox{\boldmath$\mu$}}(k_F,\tau,\phi)$ given by Eq.\,(\ref{eqn-10}),
$\mbox{\boldmath$\Omega$}_\perp(\mbox{\boldmath$k$}_\|)=\Omega_{\tau,\phi}(\mbox{\boldmath$k$}_\|)\,\hat{\mbox{\boldmath$e$}}_z$, $\Omega_{\tau,\phi}(\mbox{\boldmath$k$}_\|)=[\tau(1-\alpha^2)\pi/(1+\alpha^2)]\delta(\mbox{\boldmath$k$}_\|)$,
$\alpha=\tan\phi$, $\theta_{{\bf k}_\|}=\tan^{-1}(k_y/k_x)$,
$\mbox{\boldmath$v$}(\theta_{{\bf k}_\|})=v_F(\cos\theta_{{\bf k}_\|},\,\sin\theta_{{\bf k}_\|})$, and  $\hat{\mbox{\boldmath$e$}}_x,\,\hat{\mbox{\boldmath$e$}}_y,\,\hat{\mbox{\boldmath$e$}}_z$ are three unit coordinate vectors. In addition,
$\tilde{\mbox{\boldmath$j$}}_1(\tau,\phi,\beta_s)$ in Eq.\,(\ref{eqn-6}) represents the extrinsic non-equilibrium scattering current along the direction of a scattering angle $\beta_s$, which is different for $\tau=1$ and $-1$,
while $\mbox{\boldmath$j$}_2(\tau,\phi)$ in Eq.\,(\ref{eqn-6.2}) is the anomalous thermal-equilibrium (extrinsic) current under doping ($E_F>0$) due to Berry curvature and independent of $\beta_s$.
Furthermore, we have denoted
${\cal C}_{x,y}(k_F,\tau,\phi,\beta_s)$ as two spatial components of the vector $\mbox{\boldmath${\cal C}$}(k_F,\tau,\phi,\beta_s)=\tensor{\mbox{\boldmath$\cal T$}}_p^{-1}(k_F,\tau,\phi,\beta_s)\cdot\left\{\tensor{\mbox{\boldmath$\mu$}}(k_F,\tau,\phi)\cdot\mbox{\boldmath$E$}^\|_0\right\}$ in Eq.\,(\ref{eqn-6}).
\medskip

The elements of a conductivity tensor $\tensor{\mbox{\boldmath$\sigma$}}(\tau,\phi,\beta_s)$ can be obtained from
$\sigma_{ij}(\tau,\phi,\beta_s)=\tilde{\mbox{\boldmath$j$}}_1(\tau,\phi,\beta_s)\cdot\hat{\mbox{\boldmath$e$}}_i/(\mbox{\boldmath$E$}^\|_0\cdot\hat{\mbox{\boldmath$e$}}_j)$.
Therefore, from Eq.\,(\ref{eqn-6}), we know that the conductivity tensor depends not only on the mobility tensor, but also on the conduction-band energy dispersion and on how electrons are distributed within the conduction band.
To elucidate scattering dynamics more clearly, we study the longitudinal $j_L(\tau,\phi)$ and transverse $j_T(\tau,\phi)$ currents which flow along and perpendicular to the direction of $\beta_s$, yielding

\[
\left[\begin{array}{c}
j_L(\tau,\phi)\\
j_T(\tau,\phi)
\end{array}\right]
\equiv\int\limits_{-\pi}^{\pi} d\beta_s\,
\left[\begin{array}{c}
j_L(\tau,\phi,\beta_s)\\
j_T(\tau,\phi,\beta_s)
\end{array}\right]
=-\frac{ek_F^2\bar{\tau}_\phi(k_F,\tau)}{2\pi}\,\int\limits_{-\pi}^{\pi} d\beta_s\,
{\cal C}_x(k_F,\tau,\phi,\beta_s)
\left[\begin{array}{c}
\cos\beta_s\\
\sin\beta_s
\end{array}\right]
\]
\begin{equation}
-\frac{ek_F^2\bar{\tau}_\phi(k_F,\tau)}{2\pi}\,\int\limits_{-\pi}^{\pi} d\beta_s\,
{\cal C}_y(k_F,\tau,\phi,\beta_s)
\left[\begin{array}{c}
\sin\beta_s\\
-\cos\beta_s
\end{array}\right]\ ,
\label{eqn-6.1}
\end{equation}
where the terms containing $\cos\beta_s$ select out the diagonal elements of $\tensor{\mbox{\boldmath$\cal T$}}_p^{-1}(k_F,\tau,\phi,\beta_s)$ in Eq.\,(\ref{eqn-9}) below,
while those containing $\sin\beta_s$ keep only the off-diagonal elements of $\tensor{\mbox{\boldmath$\cal T$}}_p^{-1}(k_F,\tau,\phi,\beta_s)$.
\medskip

At low temperatures, from Eq.\,(\ref{a-10}) the thermally-averaged energy-relaxation time $\bar{\tau}_\phi(k_F,\tau)$ introduced in Eq.\,(\ref{eqn-6}) is given by

\[
\frac{1}{\bar{\tau}_\phi(k_F,\tau)}=\frac{4}{\rho_0{\cal S}}\sum_{{\bf k}_\|}\,{\cal W}^{\tau,\phi}_{\rm in}(k_\|)\Theta(k_F-|\mbox{\boldmath$k$}_\||)
\]
\begin{equation}
=\frac{4n_i}{\pi^2\hbar^2v_F\rho_0}
\int\limits_{-\pi}^{\pi} d\beta_s\,|\cos\theta|\int\limits_0^{k_F} dk_\|\,k_\|^2\left|\frac{U^\tau_0(2k_\||\cos\theta|)}{\epsilon_\phi(2k_\||\cos\theta|)}\right|^2|{\cal F}_{\tau,\phi}(k_\|,\beta_s)|^2\ ,
\label{eqn-7}
\end{equation}
which depends on both $\tau=\pm 1$ and $0\leq\phi<\pi/4$, where $|\cos\theta|=|\sin(\beta_s/2)|$, $\beta_s$ is the scattering angle,
$n_i=N_i/{\cal S}$ is the areal density of ionized impurities, and
$\epsilon_\phi(q_\|)$ is the static dielectric function obtained from Eqs.\,(\ref{e-6}) and (\ref{e-7}). Meanwhile,
the scattering form factor in Eq.\,(\ref{eqn-7}) is calculated as

\[
{\cal F}_{\tau,\phi}(k_\|,\beta_s)=\frac{1}{2}\,\sum_{\ell}\left\{(-i)^{-\tau}\tau\cos\phi\,\chi^\tau_{1,\ell}(k_\|)+\chi^\tau_{2,\ell}(k_\|)
+(-i)^\tau\tau\sin\phi\,\chi^\tau_{3,\ell}(k_\|)\right\}
\]
\[
\times\left\{(-i)^\tau\tau\cos\phi\,\chi^\tau_{1,\ell}(k_\|)\,
e^{i\tau\beta^s_{{\bf k}_\|,{\bf q}_\|}}-\chi^\tau_{2,\ell}(k_\|)
+(-i)^{-\tau}\tau\sin\phi\,\chi^\tau_{3,\ell}(k_\|)\,
e^{-i\tau\beta^s_{{\bf k}_\|,{\bf q}_\|}}\right\}
\]
\[
\equiv\kappa_0(k_\|,\phi,\tau)+\kappa_1(k_\|,\phi,\tau)\,e^{i\tau\beta_s}+\kappa_2(k_\|,\phi,\tau)\,e^{-i\tau\beta_s}+\kappa_3(k_\|,\phi,\tau)(1+e^{i\tau\beta_s})
\]
\begin{equation}
+\kappa_4(k_\|,\phi,\tau)(1+e^{-i\tau\beta_s})+\kappa_5(k_\|,\phi,\tau)\cos(\tau\beta_s)\ ,
\label{eqn-8}
\end{equation}
where $\chi^\tau_{1,\ell}(k_\|)$, $\chi^\tau_{2,\ell}(k_\|)$ and $\chi^\tau_{3,\ell}(k_\|)$ are the scattering factors defined in Eq.\,(\ref{a-20}), and six real scattering coefficients $\kappa_j$ with $j=0,\,1,\,\cdots,\,5$ can be obtained from Eq.\,(\ref{a-17}).
\medskip

In addition, from Eq.\,(\ref{b-3}) the inverse momentum-relaxation-time tensor employed in Eq.\,(\ref{eqn-6}) is microscopically calculated at low temperatures as

\[
\tensor{\mbox{\boldmath$\cal T$}}_p^{-1}(k_F,\tau,\phi)=\frac{2\pi n_i}{\rho_0}\,\left(\frac{v_F}{k_F}\right)\sum_{{\bf k}_\|,{\bf q}_\|}\,\left|U^{\tau,\phi}_{\rm im}(\mbox{\boldmath$q$}_\|,\mbox{\boldmath$k$}_\|)\right|^2\,
\delta(\varepsilon_{{\bf k}_\|}-E_F)\,\delta(\varepsilon_{{\bf k}_\|+{\bf q}_\|}-\varepsilon_{{\bf k}_\|})\,\left[\mbox{\boldmath$q$}_\|\otimes\mbox{\boldmath$q$}_\|^T\right]
\]
\[
=\frac{2n_ik^3_F}{\pi^2\hbar^2v_F\rho_0}
\int\limits_{-\pi}^{\pi} d\beta_s\,|\sin(\beta_s/2)|\sin^2(\beta_s/2)\left|\frac{U^\tau_0(2k_F|\sin(\beta_s/2)|)}{\epsilon_\phi(2k_F|\sin(\beta_s/2)|)}\right|^2
|{\cal F}_{\tau,\phi}(k_F,\beta_s)|^2
\]
\begin{equation}
\times \left[\begin{array}{cc}
\sin^2(\beta_s/2) & -\sin(\beta_s)/2\\
-\sin(\beta_s)/2 & \cos^2(\beta_s/2)
\end{array}\right]
\equiv\int\limits_{-\pi}^\pi d\beta_s\,\tensor{\mbox{\boldmath$\cal T$}}_p^{-1}(k_F,\tau,\phi,\beta_s)\ ,
\label{eqn-9}
\end{equation}
where $|{\cal F}_{\tau,\phi}(k_F,\beta_s)|^2$ has already been given by Eq.\,(\ref{eqn-8}).
It is evident from Eq.\,(\ref{eqn-9}) that the off-diagonal elements of $\tensor{\mbox{\boldmath$\cal T$}}_p^{-1}(k_F,\tau,\phi)$ become zero
after the integral has been performed with respect to $\beta_s$ from $-\pi$ to $\pi$,
while the diagonal elements of $\tensor{\mbox{\boldmath$\cal T$}}_p^{-1}(k_F,\tau,\phi)$ are nonzero and different simultaneously.
Physically, the diagonal elements of $\tensor{\mbox{\boldmath$\cal T$}}_p^{-1}(k_F,\tau,\phi,\beta_s)$ correspond to the case in which directions of the scattering force and center-of-mass momentum are parallel to each other.
The  off-diagonal elements of $\tensor{\mbox{\boldmath$\cal T$}}_p^{-1}(k_F,\tau,\phi,\beta_s)$, on the other hand, are related to a situation where the direction of the scattering force is perpendicular to that of the
center-of-mass momentum.
\medskip

Formally, by denoting the results in Eq.\,(\ref{eqn-9}) as

\begin{equation}
\displaystyle{\tensor{\mbox{\boldmath$\cal T$}}_p^{-1}(k_F,\tau,\phi)=\left[\begin{array}{cc}
b_{xx}(k_F,\tau,\phi) & 0\\
0 & b_{yy}(k_F,\tau,\phi)
\end{array}\right]}\ ,
\label{add-12}
\end{equation}
from Eqs.\,(\ref{dan-61}), (\ref{dan-63})-(\ref{dan-65}) and $\mu_{ij}(k_F,\tau,\phi)=(v_F/k_F)\,\partial K^{\tau,\phi}_i/\partial E_j$, the mobility-tensor
$\tensor{\mbox{\boldmath$\mu$}}(k_F,\tau,\phi)$ introduced in Eq.\,(\ref{eqn-6}) can easily be found as

\[
\tensor{\mbox{\boldmath$\mu$}}(k_F,\tau,\phi)=
\frac{\displaystyle{-\frac{ev_F}{\hbar k_F}}}{\displaystyle{b_{xx}(k_F,\tau,\phi)\,b_{yy}(k_F,\tau,\phi)+\left(\frac{ev_FB_z}{\hbar k_F}\right)^2}}
\]
\begin{equation}
\times\left[\begin{array}{cc}
\displaystyle{b_{yy}(k_F,\tau,\phi)} & \ \ \ \ \displaystyle{-\frac{ev_FB_z}{\hbar k_F}}\\
\displaystyle{\frac{ev_FB_z}{\hbar k_F}} & \ \ \ \ \displaystyle{b_{xx}(k_F,\tau,\phi)}
\end{array}\right]\ ,
\label{eqn-10}
\end{equation}
which depends on $\tau=\pm 1$ and $\phi$, where $\mbox{\boldmath$B$}_0^\perp=(0,0,B_z)$ introduces a normal Hall mobility (off-diagonal elements) due to broken time-reversal symmetry.
We would like to point out that the off-diagonal elements of $\tensor{\mbox{\boldmath$\cal T$}}_p^{-1}(k_F,\tau,\phi,\beta_s)$ in Eq.\,(\ref{eqn-9}) can be nonzero in principle
if an anisotropic energy dispersion $\varepsilon(\mbox{\boldmath$k$}_\|)$ contains a $k_x$ and $k_y$ crossing term,
e.g., $\varepsilon(\mbox{\boldmath$k$}_\|)\propto k_xk_y$.
\medskip

Finally, by using Eq.\,(\ref{eqn-10}), we obtain two components of the vector $\mbox{\boldmath${\cal C}$}(k_F,\tau,\phi,\beta_s)=[{\cal C}_x(k_F,\tau,\phi,\beta_s),\,{\cal C}_y(k_F,\tau,\phi,\beta_s)]
=\tensor{\mbox{\boldmath$\cal T$}}_p^{-1}(k_F,\tau,\phi,\beta_s)\cdot\left[\tensor{\mbox{\boldmath$\mu$}}(k_F,\tau,\phi)\cdot\mbox{\boldmath$E$}^\|_0\right]$
introduced in Eq.\,(\ref{eqn-6}) as

\[
{\cal C}_x(k_F,\tau,\phi,\beta_s)=-\left(\frac{ev_F}{\hbar k_F}\right)\left\{\frac{\displaystyle{b_{yy}(k_F,\tau,\phi)E_x-\left(\frac{ev_FB_z}{\hbar k_F}\right)E_y}}
{\displaystyle{b_{xx}(k_F,\tau,\phi)\,b_{yy}(k_F,\tau,\phi)+\left(\frac{qv_FB_z}{\hbar k_F}\right)^2}}\right\}\,
d_{xx}(k_F,\tau,\phi,\beta_s)
\]
\begin{equation}
-\left(\frac{ev_F}{\hbar k_F}\right)\,\left\{\frac{\displaystyle{\left(\frac{ev_FB_z}{\hbar k_F}\right)E_x+b_{xx}(k_F,\tau,\phi)E_y}}{\displaystyle{b_{xx}(k_F,\tau,\phi)\,b_{yy}(k_F,\tau,\phi)+\left(\frac{qv_FB_z}{\hbar k_F}\right)^2}}\right\}\,
d_{xy}(k_F,\tau,\phi,\beta_s)\ ,
\label{eqn-11}
\end{equation}

\[
{\cal C}_y(k_F,\tau,\phi,\beta_s)=-\left(\frac{ev_F}{\hbar k_F}\right)\,\left\{\frac{\displaystyle{\left(\frac{ev_FB_z}{\hbar k_F}\right)E_x+b_{xx}(k_F,\tau,\phi)E_y}}{\displaystyle{b_{xx}(k_F,\tau,\phi)\,b_{yy}(k_F,\tau,\phi)+\left(\frac{qv_FB_z}{\hbar k_F}\right)^2}}\right\}\,
d_{yy}(k_F,\tau,\phi,\beta_s)
\]
\begin{equation}
-\left(\frac{ev_F}{\hbar k_F}\right)\left\{\frac{\displaystyle{b_{yy}(k_F,\tau,\phi)E_x-\left(\frac{ev_FB_z}{\hbar k_F}\right)E_y}}
{\displaystyle{b_{xx}(k_F,\tau,\phi)\,b_{yy}(k_F,\tau,\phi)+\left(\frac{qv_FB_z}{\hbar k_F}\right)^2}}\right\}\,
d_{xy}(k_F,\tau,\phi,\beta_s)\ ,
\label{eqn-12}
\end{equation}
which depend on $B_z$, $\tau=\pm 1$ and Berry phase $\phi$, as well as on $\beta_s$, where $\mbox{\boldmath$E$}_0^\|=(E_x,E_y,0)$ is assumed.
In addition, $d_{xx}(k_F,\tau,\phi,\beta_s)$, $d_{yy}(k_F,\tau,\phi,\beta_s)$ and $d_{xy}(k_F,\tau,\phi,\beta_s)$ in Eq.\,(\ref{eqn-12}) are given explicitly by

\begin{eqnarray}
d_{xx}(k_F,\tau,\phi,\beta_s)&=&{\cal G}_s(k_F,\tau,\phi,\beta_s)\left(1-\cos\beta_s\right)\ ,
\nonumber
\\
d_{yy}(k_F,\tau,\phi,\beta_s)&=&{\cal G}_s(k_F,\tau,\phi,\beta_s)\left(1+\cos\beta_s\right)\ ,
\nonumber
\\
d_{xy}(k_F,\tau,\phi,\beta_s)&=&-{\cal G}_s(k_F,\tau,\phi,\beta_s)\,\sin\beta_s\ ,
\label{eqn-15}
\end{eqnarray}
where the scattering function ${\cal G}_s(k_F,\tau,\phi,\beta_s)$, which depends on $\tau,\,\phi$ and $\beta_s$, is defined as

\begin{equation}
{\cal G}_s(k_F,\tau,\phi,\beta_s)=\frac{n_ik^3_F}{4\pi^2\hbar^2v_F\rho_0}
\left|\sin^3(\beta_s/2)\right|\left|\frac{U^\tau_0(2k_F|\sin(\beta_s/2)|)}{\epsilon_\phi(2k_F|\sin(\beta_s/2)|)}\right|^2\left|{\cal F}_{\tau,\phi}(k_F,\beta_s)\right|^2\ .
\label{eqn-16}
\end{equation}
In Eqs.\,(\ref{eqn-11}) and (\ref{eqn-12}), the terms containing $d_{xy}(k_F,\tau,\phi,\beta_s)$ represent the contributions to skew scattering.

\section{Numerical Results and Discussions}
\label{sec-3}

In our numerical calculations, we take: $v_F=10^8\,$cm/s, $\rho_0=5\times 10^{11}\,$cm$^{-2}$, $k_F=\sqrt{\pi\rho_0}$, $E_F=\hbar v_Fk_F$, $n_i=2.5\times 10^{11}\,$cm$^{-2}$, $Z^*=2$, $\epsilon_r=13$, $\Lambda_\|=100\,$\AA,
$V_0/E_F=1.4$, $k_Fr_0=7$, $E_x=0.5\,k$V/cm, and $E_y=0$.
The other parameters, such as, $\phi$, $\tau$ and $B_z$, will be directly given in figure captions.
\medskip

\begin{figure}
\centering
\includegraphics[width=0.85\textwidth]{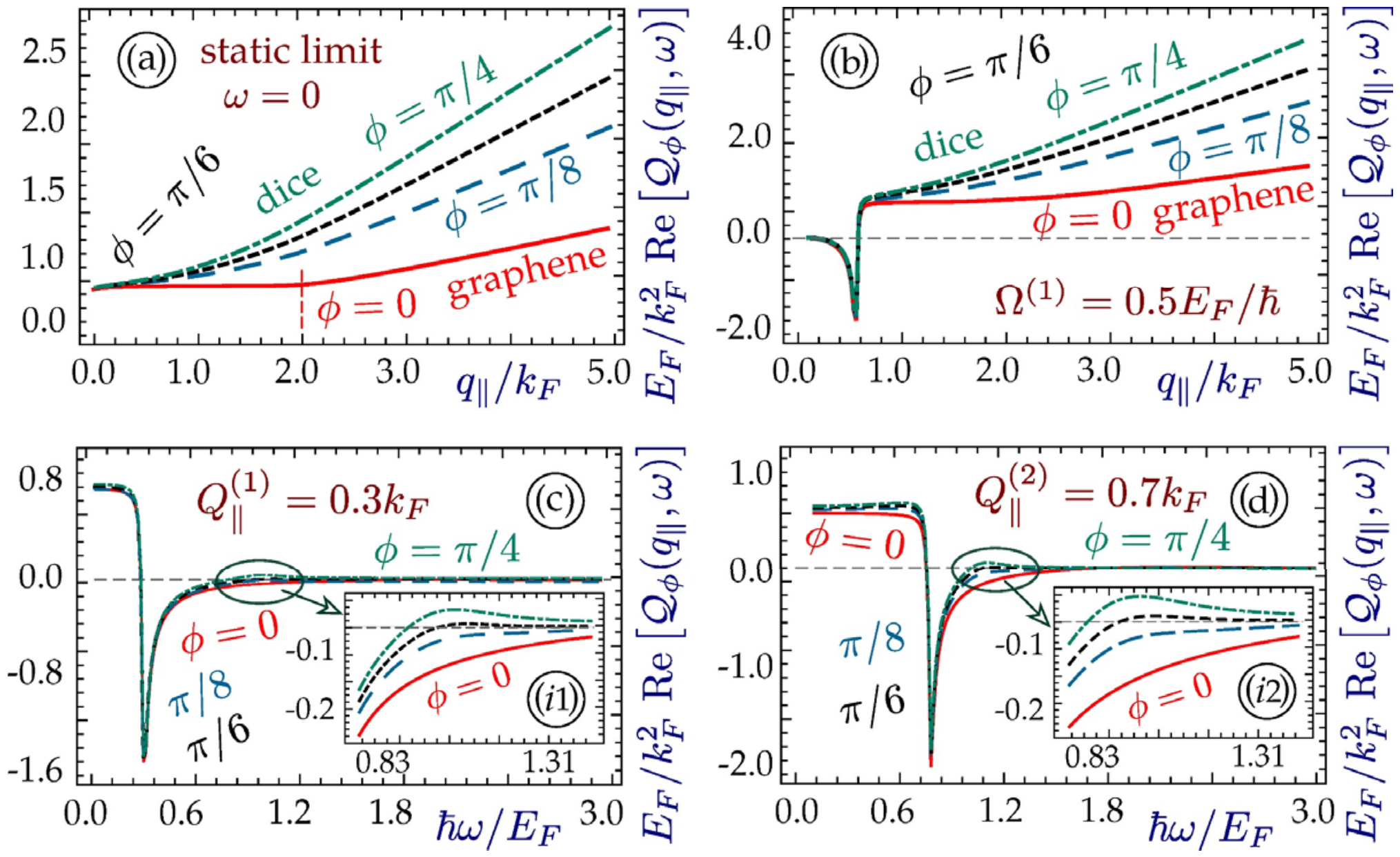}
\caption{Calculated real part of the polarization function
${\rm Re}[{\cal Q}_\phi(\mbox{\boldmath$q$}_\|,\omega)]$ from Eq.\,(\ref{e-2}) with $\phi=\pi/4$ (dice, green), $\pi/6$ (black), $\pi/8$ (blue), and $0$ (graphene, red) as a function of $q_\|$ at $\hbar\omega=0$ ($a$)
and $\hbar\omega/E_F=0.5$ ($b$), as well as a function of $\hbar\omega$ at $q_\|/k_F=0.3$ ($c$) and $q_\|/k_F=0.7$ ($d$).
Here, the unit of $(k^2_F/E_F)$ has been used for scaling ${\cal Q}_\phi(\mbox{\boldmath$q$}_\|,\omega)$ in Eq.\,(\ref{e-2}).}
\label{fig2}
\end{figure}

Using Eq.\,(\ref{e-2}), we have shown in Fig.\,\ref{fig2} the real part of the polarization function ${\rm Re}[{\cal Q}_\phi(q_\|,\omega)]$
as a function of $q_\|$ at $\hbar\omega/E_F=0$ ($a$) and $0.5$ ($b$) and as a function of $\hbar\omega$ at $q_\|/k_F=0.3$ ($c$)
and $0.7$ ($d$). We know from Fig.\,\ref{fig2}($a$) that all results with different $\phi$ approach a finite constant as $q_\|\to 0$ in the static limit ($\omega=0$), including graphene with $\phi=0$ within the whole region of $q_\|/k_F\leq 2$.
However, they increase significantly with $q_\|$ as $q_\|/k_F>2$ and become strongly $\phi$ dependent.
These features in Fig.\,\ref{fig2}($a$) change completely for $\hbar\omega/E_F=0.5$, as shown in Fig.\,\ref{fig2}($b$), where ${\rm Re}[{\cal Q}_\phi(q_\|,\omega)]=0$ (i.e., no screening) at $q_\|=0$ for all values of $\phi$.
Figures\ \ref{fig2}($c$) and \ref{fig2}($d$) display ${\rm Re}[{\cal Q}_\phi(q_\|,\omega)]$ as a function of $\hbar\omega$ at $q_\|/k_F=0.3$ ($c$) and $0.7$ ($d$),
where a sharp and nearly $\phi$-independent negative peak shifts up rapidly in frequency as $q_\|$ increases. Moreover, a series of intersections with the thin dashed line (i.e., ${\rm Re}[{\cal Q}_\phi(q_\|,\omega)]=0$)
is seen in the two insets in Figs.\,\ref{fig2}($c$) and \ref{fig2}($d$).
This highlights a sign switch of ${\rm Re}[{\cal Q}_\phi(q_\|,\omega)]$ and implies the existence of a set of $\phi$-dependent plasmon resonances
determined from ${\rm Re}[{\cal Q}_\phi(q_\|,\omega)]=-\alpha q_\|/k_F$ with $\alpha=2\epsilon_0\epsilon_r\hbar v_F/e^2$ on the right-hand-side shoulder of this negative peak.
\medskip

\begin{figure}
\centering
\includegraphics[width=0.85\textwidth]{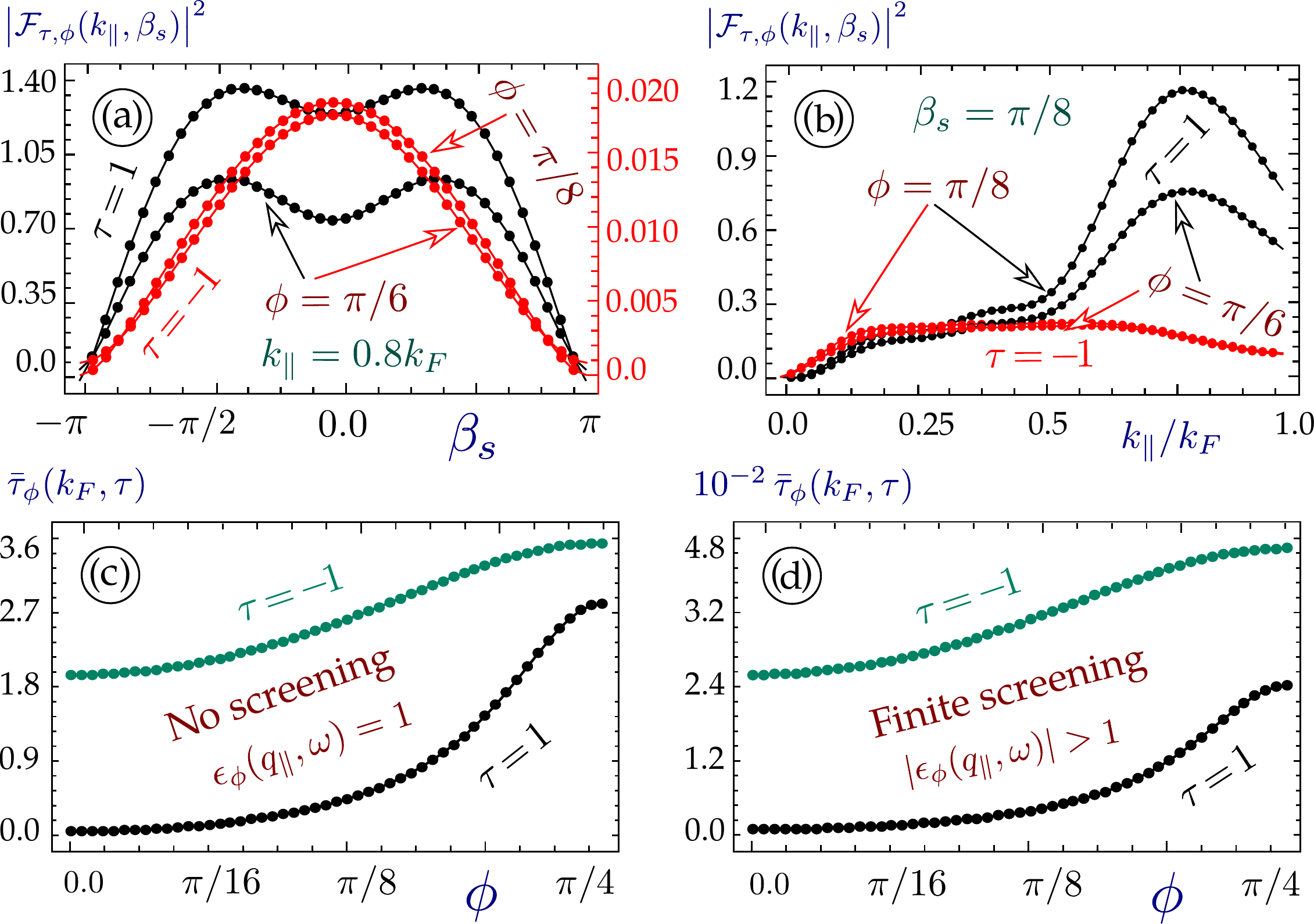}
\caption{Calculated square of the dimensionless form factor $|{\cal F}_{\tau,\phi}(k_\|,\beta_s)|^2$ from Eq.\,(\ref{eqn-8}) with $\phi=\pi/6$ and $\pi/8$
as a function of $\beta_s$ at $k_\|/k_F=0.8$ ($a$) and as a function of $k_\|$ at $\beta_s=\pi/8$ ($b$) for $\tau=1$ (black) and $\tau=-1$ (red);	
as well as thermally-averaged energy-relaxation time $\bar{\tau}_{\phi}(k_F,\tau)$ calculated from Eq.\,(\ref{eqn-7}) as a function of $\phi$ for $\tau=1$ (black) and $\tau=-1$ (green) under both unscreened ($c$) and screened ($d$) conditions.
Here, the unit of $(\pi^2\hbar/4E_F)$ has been used for scaling $\bar{\tau}_{\phi}(k_F,\tau)$.}
\label{fig3}
\end{figure}

We present the calculated square of the form factor $|{\cal F}_{\tau,\phi}(k_\|,\beta_s)|^2$ in Fig.\,\ref{fig3} for $\tau=\pm 1$
by using Eq.\,(\ref{eqn-8}) as a function of the scattering angle $\beta_s$ at $k_\|/k_F=0.8$ ($a$) and as a function of the wave number $k_\|/k_F$ at $\beta_s=\pi/8$ ($b$)
with $\phi=\pi/8$ and $\pi/6$. From Fig.\,\ref{fig3}($a$), we find either a single peak or double peaks with respect to $\beta_s$
for $\tau=1$ (black, left-scale) or $\tau=-1$ (red, right-scale), respectively.
This valley-dependent behavior of $|{\cal F}_{\tau,\phi}(k_\|,\beta_s)|^2$ is attributed to different barrier-like (trap-like) impurity scattering for $\tau=1$ ($\tau=-1$), and
the latter only acquires a weak strength. Moreover, we find from Fig.\,\ref{fig3}($b$) that significant difference in $|{\cal F}_{\tau,\phi}(k_\|,\beta_s)|^2$ for $\tau=\pm 1$ exists only
for large $k_\|$ values ($k_\|/k_F\geq 0.5$). This valley dependence of $|{\cal F}_{\tau,\phi}(k_\|,\beta_s)|^2$ has a profound influence on the energy-relaxation time $\bar{\tau}_{\phi}(k_F,\tau)$, as demonstrated by Fig.\,\ref{fig3}($c$) and \ref{fig3}($d$),
where $\bar{\tau}_{\phi}(k_F,\tau)$ calculated from Eq.\,(\ref{eqn-7}) is displayed as a function of Berry phase $\phi$ for $\tau=1$ and $-1$ under both unscreened ($c$) with $\epsilon_\phi(q_\|,\omega)\equiv 1$
and screened ($d$) conditions. By comparing Figs.\,\ref{fig3}($c$) with \ref{fig3}($d$), it is apparent that the strength of impurity scattering can be overestimated by
almost two orders of magnitude if the many-body screening effect has been neglected.
Meanwhile, $\bar{\tau}_{\phi}(k_F,\tau)$ increases monotonically with $\phi$, and it becomes larger for $\tau=-1$, in comparison with that for $\tau=1$, due to a weaker trap-like impurity scattering of electrons.
Furthermore, the difference in $\bar{\tau}_{\phi}(k_F,\tau)$ under screening for two valleys remains unchanged for all values of $\phi$.
\medskip

\begin{figure}
\centering
\includegraphics[width=0.85\textwidth]{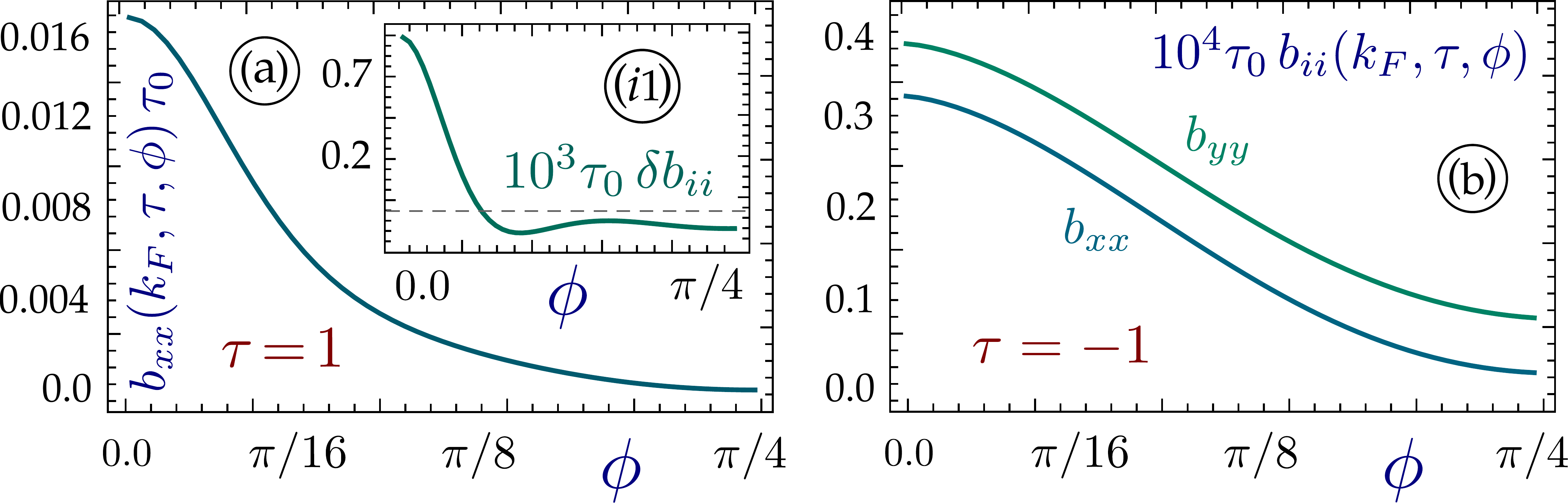}
\caption{Calculated diagonal elements $b_{xx}(k_F,\tau,\phi)$ for $\tau=1$ ($a$) and both $b_{xx}(k_F,\tau,\phi)$ and $b_{yy}(k_F,\tau,\phi)$ for $\tau=-1$ ($b$)
of the inverse momentum-relaxation-time tensor $\tensor{\mbox{\boldmath$\cal T$}}_p^{-1}(k_F,\tau,\phi)$ in Eq.\,(\ref{add-12})
as functions of $\phi$, where the difference $\delta b\equiv b_{xx}(k_F,\tau,\phi)-b_{yy}(k_F,\tau,\phi)$ for $\tau=1$ is also presented in the inset ($i1$)
and the dashed line corresponds to $\delta b=0$ to highlight its sign switching.
Here, the unit of $1/\tau_0=4E_F/\pi^2\hbar$ has been used for scaling $b_{xx}(k_F,\tau,\phi)$ and $b_{yy}(k_F,\tau,\phi)$.}
\label{fig4}
\end{figure}

The calculated two diagonal elements, $b_{xx}(k_F,\tau,\phi)$ and $b_{yy}(k_F,\tau,\phi)$, of the inverse momentum-relaxation-time tensor $\tensor{\mbox{\boldmath$\cal T$}}_p^{-1}(k_F,\tau,\phi)$ in Eq.\,(\ref{add-12})
are presented in Figs.\,\ref{fig4}($a$) and \ref{fig4}($b$) as a function of $\phi$ for $\tau=1$ and $-1$, respectively.
We first notice from Fig.\,\ref{fig4}($b$) that $b_{xx}(k_F,\tau,\phi)$ is lower than $b_{yy}(k_F,\tau,\phi)$,
but both of them decrease monotonically with $\phi$ in a similar way. Also, we would like to point out that
the rate difference $\delta b\equiv b_{xx}(k_F,\tau,\phi)-b_{yy}(k_F,\tau,\phi)$, as shown by the inset in Fig.\,\ref{fig4}($a$),
decreases with $\phi$ initially but switches to negative and saturates afterwards for large $\phi$ values.
Contrary to the result in Fig.\,\ref{fig4}($b$), we find $b_{xx}(k_F,\tau,\phi)>b_{yy}(k_F,\tau,\phi)$ in Fig.\,\ref{fig4}($a$) before the sign switch of $\delta b$.
Moreover, $b_{xx}(k_F,\tau,\phi)$ and $b_{yy}(k_F,\tau,\phi)$ in Fig.\,\ref{fig4}($a$) are more than two orders of magnitude higher than those in Fig.\,\ref{fig4}($b$),
implying an enhanced momentum-dissipation rate for electrons at the $\tau=1$ valley due to much larger $|{\cal F}_{\tau,\phi}(k_\|,\beta_s)|^2$
for $\tau=1$ and $k_\|=k_F$ in Fig.\,\ref{fig3}($b$).
\medskip

\begin{figure}
\centering
\includegraphics[width=0.85\textwidth]{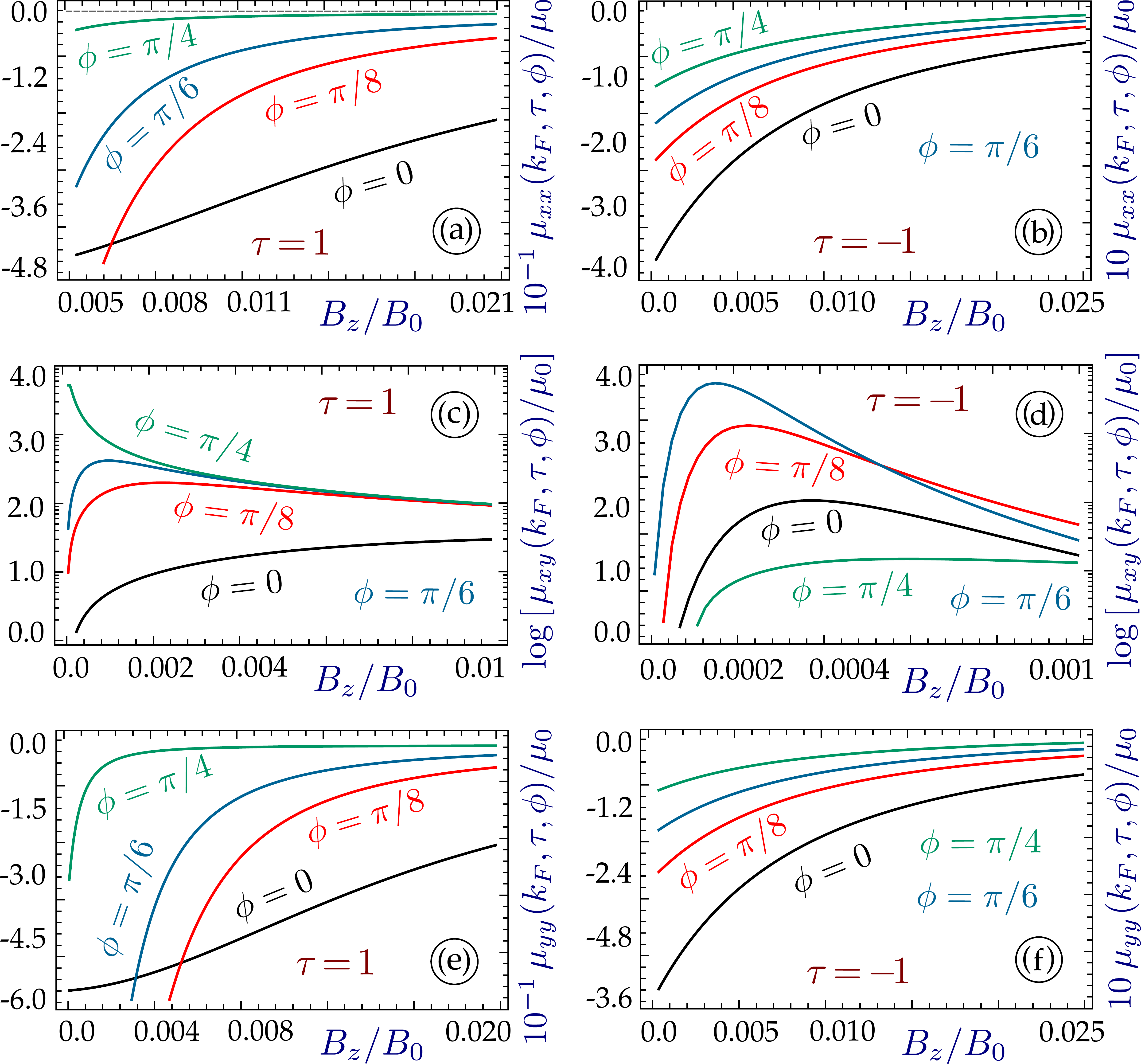}
\caption{Calculated diagonal elements $\mu_{xx}(k_F,\tau,\phi)$ ($a$),($b$) and $\mu_{yy}(k_F,\tau,\phi)$ ($e$),($f$), as well as off-diagonal element $\mu_{xy}(k_F,\tau,\phi)$ in logarithm scale ($c$),($d$),
of the mobility tensor $\tensor{\mbox{\boldmath$\mu$}}(k_F,\tau,\phi)$ given by Eq.\,(\ref{eqn-10})
as a function of $B_z$ with $\phi=\pi/4$ (green), $\phi=\pi/6$ (blue), $\phi=\pi/8$ (red) and $\phi=0$ (black) for $\tau=1$ ($a$),\,($c$),\,($e$) and $\tau=-1$ ($b$),\,($d$),\,($f$).
Here, $\mu_0=4e/\pi^2\hbar k_F^2$ has been used for scaling all elements of $\tensor{\mbox{\boldmath$\mu$}}(k_F,\tau,\phi)$ and $B_0=\hbar k_F^2/e$.}
\label{fig5}
\end{figure}

In Fig.\,\ref{fig5} we exhibit two diagonal elements, $\mu_{xx}(k_F,\tau,\phi)$ ($a$)-($b$) and $\mu_{yy}(k_F,\tau,\phi)$ ($e$)-($f$),
as well as the off-diagonal element, $\mu_{xy}(k_F,\tau,\phi)$ ($c$)-($d$), of the mobility tensor $\tensor{\mbox{\boldmath$\mu$}}(k_F,\tau,\phi)$
in Eq.\,(\ref{eqn-10}) as a function of magnetic field $B_z$ for four different Berry phases and $\tau=\pm 1$.
By comparing Figs.\,\ref{fig5}($a$), \ref{fig5}($c$) and \ref{fig5}($e$) for $\tau=1$ with Figs.\,\ref{fig5}($b$), \ref{fig5}($d$)
and \ref{fig5}($f$) for $\tau=-1$,
we discover significant difference between their dependence and magnitudes due to two orders of magnitude change in $b_{xx}(k_F,\tau,\phi)$ and
$b_{yy}(k_F,\tau,\phi)$ in Fig.\,\ref{fig4} for $\tau=1$ and $-1$. The longitudinal mobilities $\mu_{xx}(k_F,\tau,\phi)$
and $\mu_{yy}(k_F,\tau,\phi)$, related to back scattering of electrons,
are somewhat suppressed not only by increasing the Lorentz force (or increasing $B_z$) in position space due to cyclotron motion, but also
by decreasing the Berry force (or decreasing Berry curvature $\Omega_s^{\tau,\phi}(\mbox{\boldmath$k$}_\|)$) in momentum space.
For high $B_z$, we arrive at $\mu_{xx},\,\mu_{yy}\sim 1/B_z^2$, corresponding to a classical limit.
In addition, the transverse mobility $\mu_{xy}(k_F,\tau,\phi)$, connected to skew scattering of electrons,
also decreases with reduced Berry force in momentum space at low $B_z$, where an initial sharp increase
(logarithm scale in Figs.\,\ref{fig5}($c$)-\ref{fig5}($d$))
of $\mu_{xy}(k_F,\tau,\phi)$ is found slightly above $B_z=0$ but it quickly changes to decreasing with $B_z$ until a classical limit,
i.e., $\mu_{xy}\sim 1/B_z$, is reached in the strong-field limit.
\medskip

\begin{figure}
\centering
\includegraphics[width=0.85\textwidth]{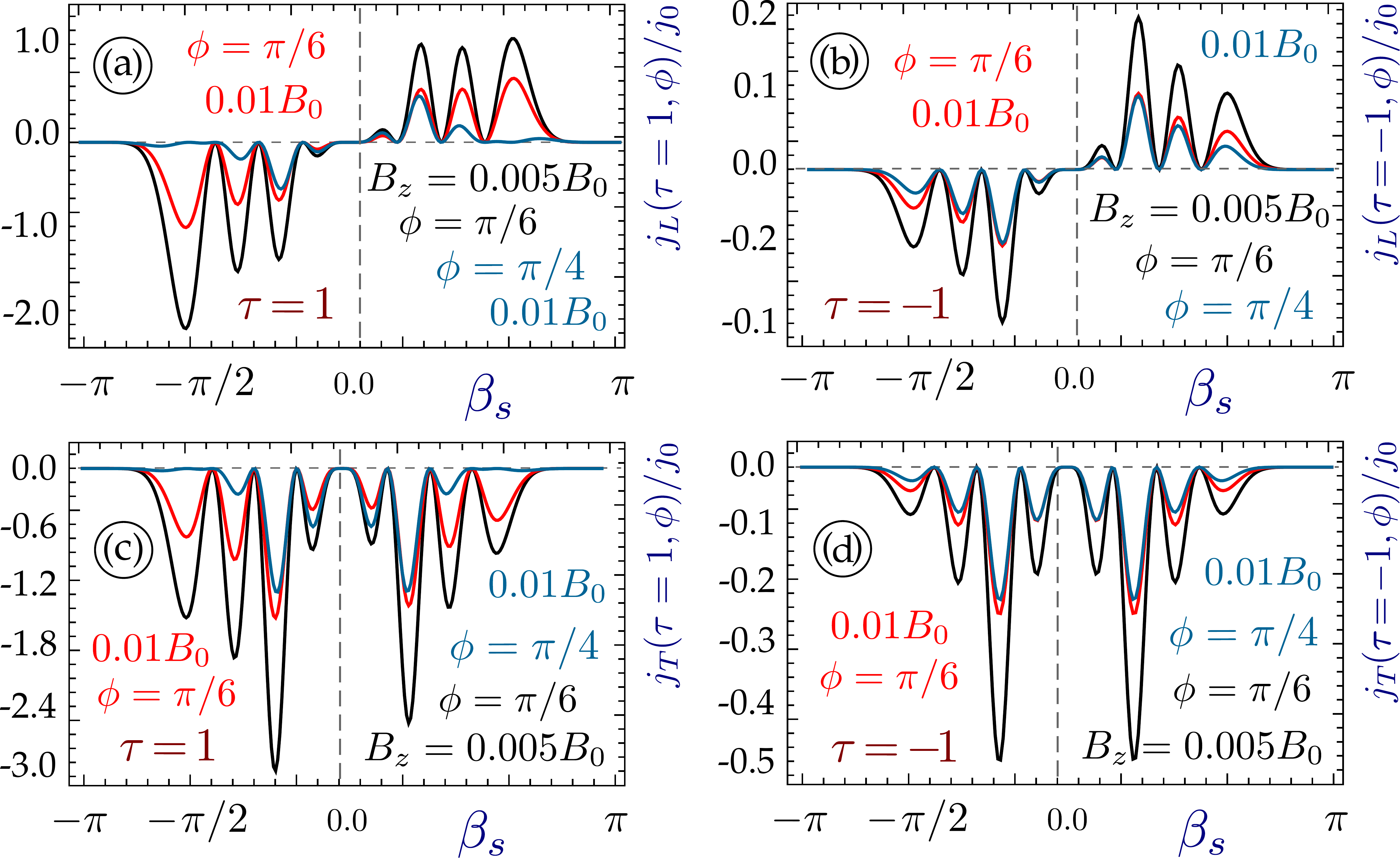}
\caption{Calculated integrands of longitudinal $j_L(\tau,\phi)$ ($a$)-($b$) and transverse $j_T(\tau,\phi)$ ($c$)-($d$) scattering currents from Eq.\,(\ref{eqn-6.1})
as a function of $\beta_s\in[-\pi,\pi]$ with $\phi=\pi/4,\,B_z/B_0=0.01$ (blue), $\phi=\pi/6,\,B_z/B_0=0.01$ (red) and $\phi=\pi/6,\,B_z/B_0=0.005$ (black)
for $\tau=1$ ($a$),\,($c$) and $\tau=-1$ ($b$),\,($d$). Here, the unit of $j_0=n_iev_F$ has been used for scaling both $j_L(\tau,\phi)$ and $j_T(\tau,\phi)$ and $B_0$ is given in Fig.\,\ref{fig5}.}
\label{fig6}
\end{figure}

After presenting a full calculation of physical parameters of $\alpha$-$T_3$ lattices in Figs.\,\ref{fig2}-\ref{fig5}, we turn to discussions on
valley-dependent electrical responses, i.e., gVHE on directly-measurable sheet current density. To clearly reveal valley scattering dynamics,
we show in Fig.\,\ref{fig6} the scattering-angle ($\beta_s$) distributions
of longitudinal $j_L(\tau,\phi,\beta_s)$ ($a$)-($b$) and transverse $j_T(\tau,\phi,\beta_s)$ ($c$)-($d$) currents given by
Eq.\,(\ref{eqn-6.1}) with various Berry phases $\phi$ and $B_z$ for $\tau=1$ ($a$),\,($c$) and $\tau=-1$ ($b$),\,($d$).
From Figs.\,\ref{fig6}($a$) and \ref{fig6}($b$) we see a triplet peak in $j_L(\tau,\phi)$ with opposite signs for $\beta_s>0$ and $\beta_s<0$. Much more interestingly,
we always find one backward plus one forward near-vertical (near-horizontal) scattering of electrons from two different valley impurities, characterized by $\tau=1$ ($\tau=-1$) here.
As expected, $j_L(\tau,\phi)$ for $\tau=1$ is one order of magnitude higher than that for $\tau=-1$ because of a larger mobility for the former.
The increase of $B_z$ significantly reduces
$j_L(\tau,\phi)$ at $\phi=\pi/6$ for both $\tau=\pm 1$ (black and red) due to cyclotron motion. Meanwhile, the increase of Berry phase $\phi$ further reduces $j_L(\tau,\phi)$ at $B_z/B_0=0.01$
for both $\tau=\pm 1$ (red and blue) due to decreasing Berry force.
Furthermore, the negative triplet peak is always present for $j_T(\tau,\phi)$ in both $\beta_s>0$ and $\beta_s<0$ regions,
as shown in Figs.\,\ref{fig6}($c$) and \ref{fig6}($d$).
Here, $j_T(\tau,\phi)$ exhibits the same dependence as for the triplet peak in $j_L(\tau,\phi)$ on $B_z$ and $\phi$.
In this case, however, one always finds a counter-clockwise tangential current $j_T(\tau,\phi)$ for dominant near-horizontal forward- and backward-scattering of electrons with an impurity at both valleys.
\medskip

\begin{figure}
\centering
\includegraphics[width=0.85\textwidth]{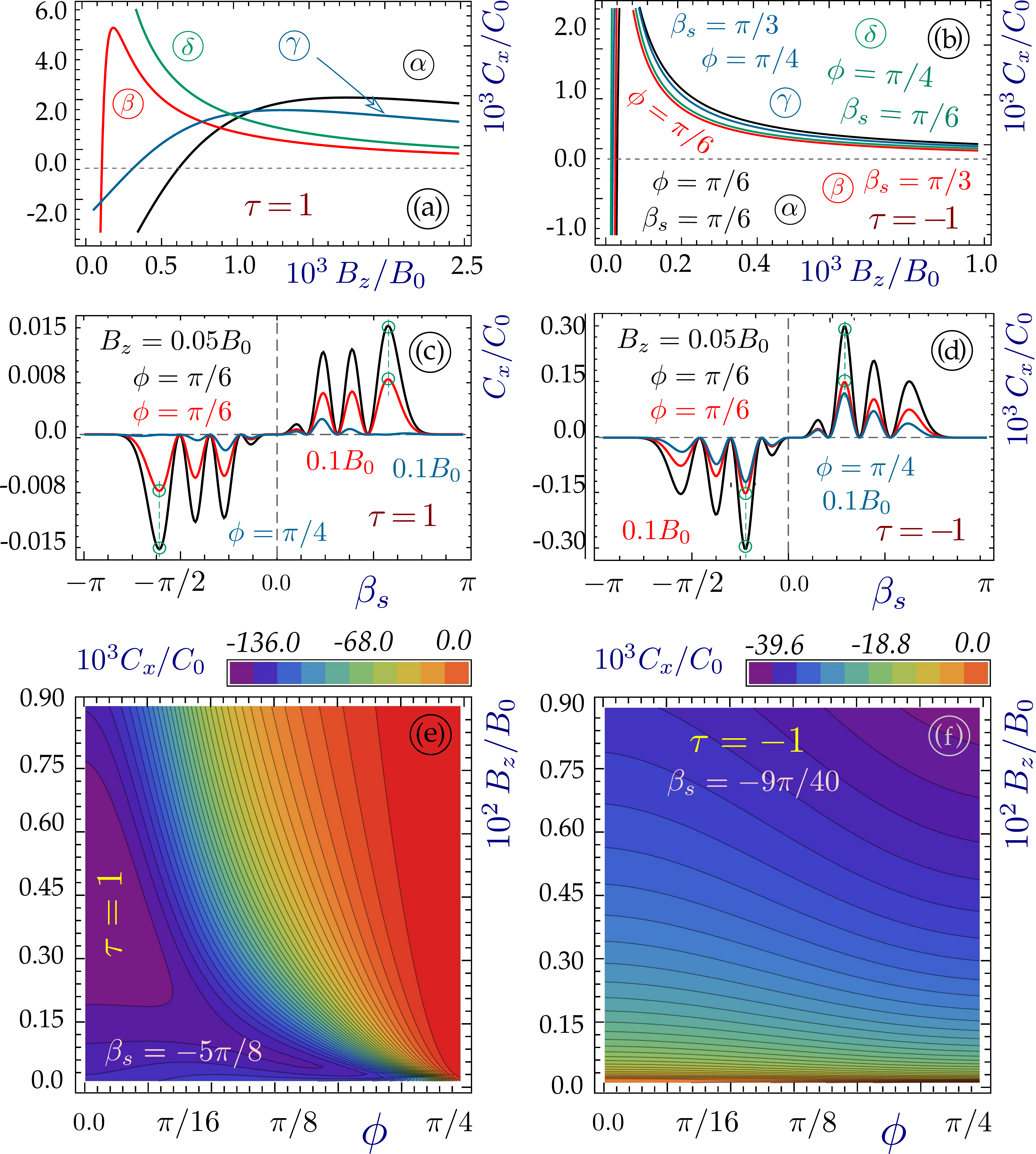}
\caption{($a$)-($d$) back-scattering current-distribution component $C_x(k_F,\tau,\phi,\beta_s)$ from Eq.\,(\ref{eqn-11})
as a function of $B_z$ ($a$),($b$) with $\phi=\pi/4,\,\beta_s=\pi/6$ (green), $\phi=\pi/6,\,\beta_s=\pi/6$ (black), $\phi=\pi/4,\,\beta_s=\pi/3$ (blue) and $\phi=\pi/6,\,\beta_s=\pi/3$ (red)
for $\tau=1$ ($a$) and $\tau=-1$ ($b$),
as well as a function of $\beta_s$ with $\phi=\pi/6,\,B_z/B_0=0.05$ (black), $\phi=\pi/6,\,B_z/B_0=0.1$ (red) and $\phi=\pi/4,\,B_z/B_0=0.1$ (blue)
for $\tau=1$ ($c$) and $\tau=-1$ ($d$);
2D contour plots of $C_x(k_F,\tau,\phi,\beta_s)$ ($e$),($f$) as a function of both
$\phi$ and $B_z$ for $\beta_s=-5\pi/8$ and $\tau=1$ ($e$) and for $\beta_s=-9\pi/40$ and $\tau=-1$ ($f$).
Here, two green circles in ($c$),\,($d$) indicate large back-scattering current peaks at $\beta_s\approx-5\pi/8$ ($\beta_s\approx-9\pi/40$) for $\tau=1$ ($\tau=-1$), respectively.
In addition, the unit of $C_0=4k_Fv_F^2/\pi^2$ has been used for scaling $C_x(k_F,\tau,\phi,\beta_s)$ and $B_0$ is given in Fig.\,\ref{fig5}.}
\label{fig7}
\end{figure}

In order to gain a better physics picture about the valley-dependent triplet peak
of the longitudinal scattering currents in Figs.\,\ref{fig6}($a$) and \ref{fig6}($b$),
we present in Fig.\,\ref{fig7} the back-scattering current-distribution component $C_x(k_F,\tau,\phi,\beta_s)$ from Eq.\,(\ref{eqn-11}) as a function of either $B_z$
or $\beta_s$, as well as 2D contour plots of $C_x(k_F,\tau,\phi,\beta_s)$ as a function of both $\phi$ and $B_z$ for $\tau=1$ ($\tau=-1$) and $\beta_s=-5\pi/8$ ($\beta_s=-9\pi/40$), respectively.
We find from Figs.\,\ref{fig7}($a$) and \ref{fig7}($b$) that for all cases $C_x(k_F,\tau,\phi,\beta_s)$ is initially increased but subsequently
reduced by a magnetic field for both $\tau=\pm 1$. Increasing $\phi$ from $\pi/6$ (black) to $\pi/4$ (green) at fixed $\beta_s=\pi/6$ can switch the sign of (reduce)
$C_x(k_F,\tau,\phi,\beta_s)$ for $\tau=1$ ($\tau=-1$) at low $B_z$. An opposite situation occurs at $\beta_s=\pi/3$, but experiences a smaller change for $\tau=1$.
On the other hand, from Figs.\,\ref{fig7}($c$) and \ref{fig7}($d$) we see one backward plus one forward weak near-vertical (very strong near-horizontal) scattering for $\tau=1$ ($\tau=-1$), respectively,
with similar features as those found in Figs.\,\ref{fig6}($a$) and \ref{fig6}($b$) for their dependence on $B_z$ and $\phi$.
The contour plot at $\beta_s=-5\pi/8$ and $\tau=1$ in Fig.\,\ref{fig7}($e$) displays an ``island'' in $C_x(k_F,\tau,\phi,\beta_s)$
at the left side of this panel associated with low $\phi$ and intermediate $B_z$ values.
For $\tau=-1$ and $\beta_s=-9\pi/40$ in Fig.\,\ref{fig7}($f$), however, only a negative peak at bottom is found for very low $B_z$.
Such distinctive features in Figs.\,\ref{fig7}($e$) and \ref{fig7}($f$) present a clear proof to the existence of gVHE in the current system.
\medskip

\begin{figure}
\centering
\includegraphics[width=0.65\textwidth]{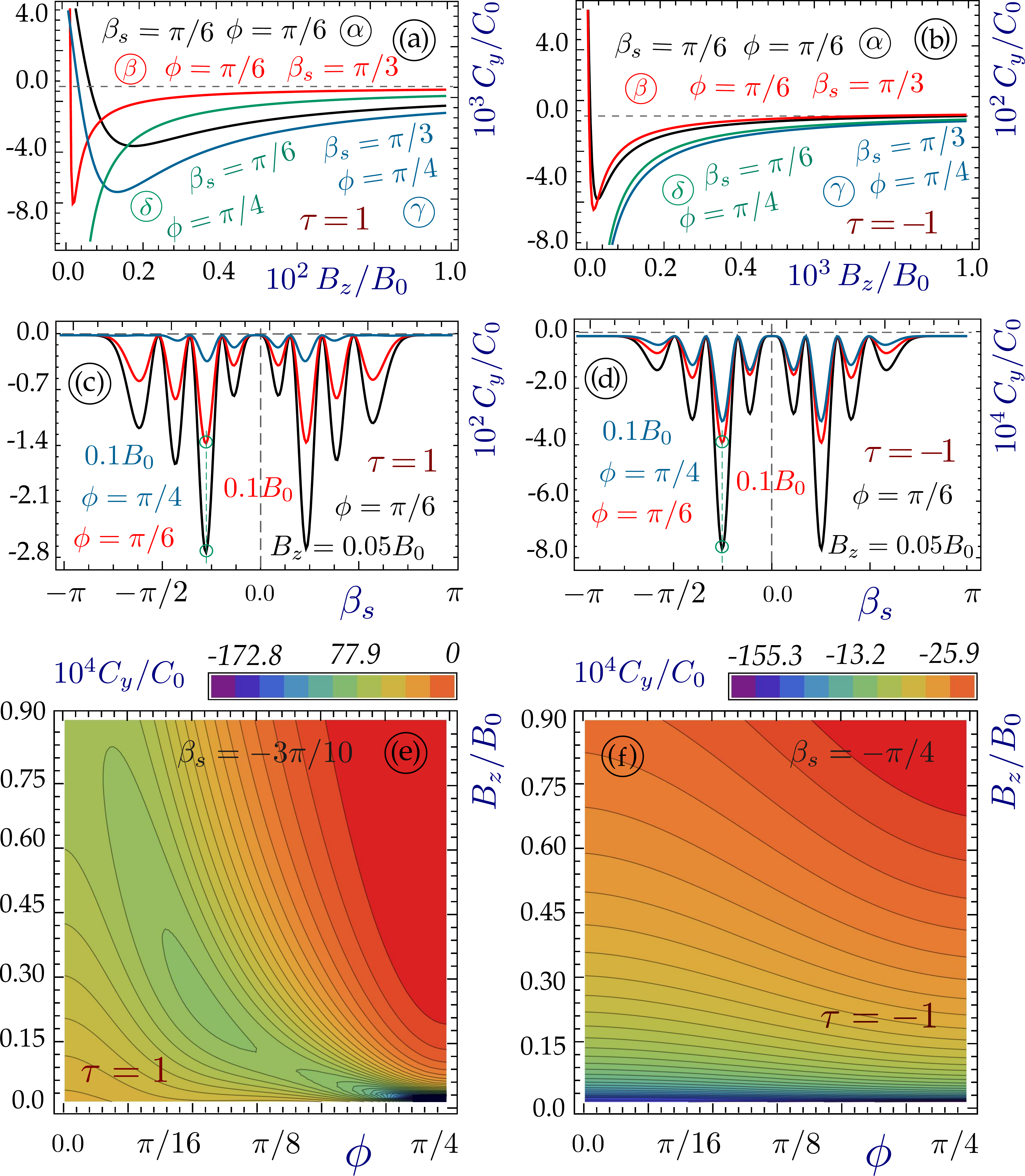}
\caption{($a$)-($d$) skew-scattering current-distribution component $C_y(k_F,\tau,\phi,\beta_s)$ from Eq.\,(\ref{eqn-12})
as a function of $B_z$ ($a$)-($b$) with $\phi=\pi/4,\,\beta_s=\pi/6$ (green), $\phi=\pi/6,\,\beta_s=\pi/6$ (black), $\phi=\pi/4,\,\beta_s=\pi/3$ (blue) and $\phi=\pi/6,\,\beta_s=\pi/3$ (red)
for $\tau=1$ ($a$) and $\tau=-1$ ($b$),
as well as a function of $\beta_s$ with $\phi=\pi/6,\,B_z/B_0=0.05$ (black), $\phi=\pi/6,\,B_z/B_0=0.1$ (red) and $\phi=\pi/4,\,B_z/B_0=0.1$ (blue)
for $\tau=1$ ($c$) and $\tau=-1$ ($d$);
($e$)-($f$) 2D contour plots of $C_y(k_F,\tau,\phi,\beta_s)$ as a function of both
$\phi$ and $B_z$ for $\beta_s=-3\pi/10$ and $\tau=1$ ($e$) and for $\beta_s=-\pi/4$ and $\tau=-1$ ($f$).
Here, two green circles in ($c$),\,($d$) indicate large skew-current peaks at $\beta_s\approx-3\pi/10$ ($\beta_s=-\pi/4$) for $\tau=1$ ($\tau=-1$), respectively.
In addition, $C_0$ and $B_0$ are given in Figs.\,\ref{fig7} and \ref{fig5}, respectively.}
\label{fig8}
\end{figure}

We also plot in Fig.\,\ref{fig8} the skew-scattering current-distribution component $C_y(k_F,\tau,\phi,\beta_s)$ from Eq.\,(\ref{eqn-12}) as a function of $B_z$
and $\beta_s$, as well as 2D contour plots of $C_y(k_F,\tau,\phi,\beta_s)$ as a function of both $\phi$ and $B_z$ for $\tau=1$ ($\tau=-1$) and $\beta_s=-3\pi/10$ ($\beta_s=-\pi/4$), respectively.
We observe from Figs.\,\ref{fig8}($a$) and \ref{fig8}($b$) that for all cases $C_y(k_F,\tau,\phi,\beta_s)$ initially switches sign slow (fast) but subsequently
decreases with $B_z$ for $\tau=1$ ($\tau=-1$), different from the results in Figs.\,\ref{fig7}($a$) and \ref{fig7}($b$).
Increasing $\phi$ from $\pi/6$ (red) to $\pi/4$ (blue) at $\beta_s=\pi/3$ will reduce (enhance)
$C_y(k_F,\tau,\phi,\beta_s)$ for $\tau=1$ ($\tau=-1$) at very low $B_z$. However, $C_y(k_F,\tau,\phi,\beta_s)$ is always enhanced with $\phi$ for another scattering angle at $\beta_s=\pi/6$ with a bigger variation for $\tau=-1$.
From Figs.\,\ref{fig8}($c$) and \ref{fig8}($d$), we only see a strong (weak) sharp negative triplet skew-scattering peak in the full region of $\beta_s$ with similar features as those found in Figs.\,\ref{fig6}($c$) and \ref{fig6}($d$)
for their dependence on $B_z$ and $\phi$ at $\tau=1$ ($\tau=-1$). This leads to upward currents for both near-vertical (near-horizontal)
forward- and backward-scattering at $\tau=1$
($\tau=-1$), respectively.
The contour plot with $\beta_s=-3\pi/10$ and $\tau=1$ in Fig.\,\ref{fig8}($e$) again reveals a unique strong negative peak in $C_y(k_F,\tau,\phi,\beta_s)$
at the lower-right corner of this panel. For $\tau=-1$ and $\beta_s=-\pi/4$ in Fig.\,\ref{fig8}($f$), on the other hand,
only one negative peak at bottom is seen for very small $B_z$, similar to that in Fig.\,\ref{fig7}($f$).
\medskip

\begin{figure}
\centering
\includegraphics[width=0.85\textwidth]{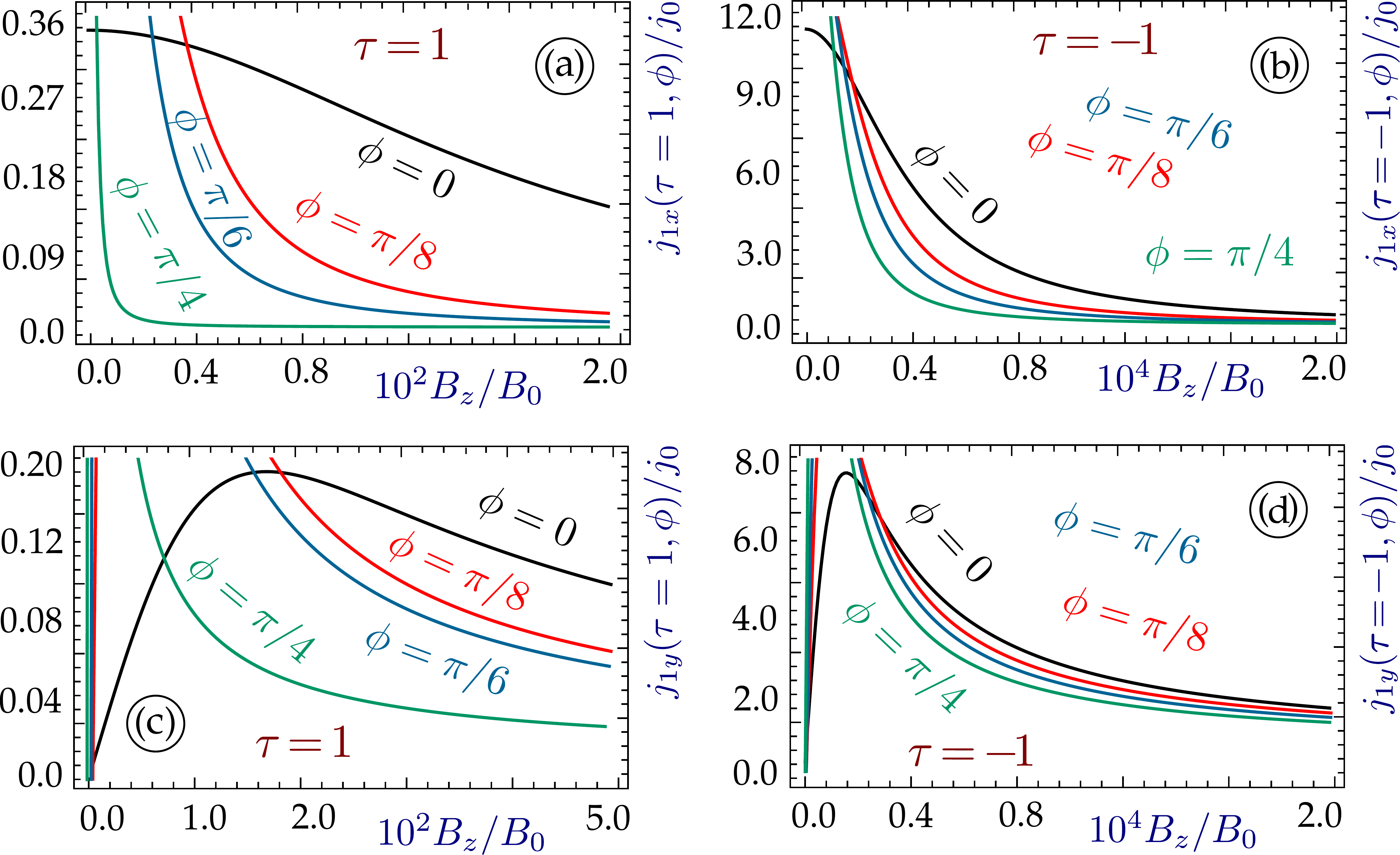}
\caption{Calculated non-equilibrium total back-scattering current $j_{1x}(\tau,\phi)$ ($a$)-($b$) and total skew-scattering current $j_{1y}(\tau,\phi)$ ($c$)-($d$) from Eq.\,(\ref{eqn-6})
as a function of $B_z$ with $\phi=\pi/4$ (green), $\phi=\pi/6$ (blue), $\phi=\pi/8$ (red) and $\phi=0$ (black)
for $\tau=1$ ($a$),\,($c$) and $\tau=-1$ ($b$),\,($d$). Here $B_0$ and $j_0$ are given in Figs.\,\ref{fig5} and \ref{fig6}, respectively.}
\label{fig9}
\end{figure}

For a comparison with experimentally measurable currents, we display in Fig.\,\ref{fig9} the
calculated total back-scattering current $j_{1x}(\tau,\phi)$ in ($a$)-($b$), as well as
total skew-scattering current $j_{1y}(\tau,\phi)$ in ($c$)-($d$), from Eq.\,(\ref{eqn-6}) as a function of $B_z$ with various phases $\phi$
for $\tau=1$ ($a$), ($c$) and $\tau=-1$ ($b$), ($d$).
From  Figs.\,\ref{fig9}($a$) and \ref{fig9}($b$), we see a slow (fast) monotonic decrease of $j_{1x}(\tau,\phi)$ with
increasing $B_z$ in the scale of $\sim 1/B_z^2$ for $\tau=1$ ($\tau=-1$) due to cyclotron motion.
Such different behaviors are attributed to lower (higher) mobility at the $\tau=1$ ($\tau=-1$) valley.
However, increasing $\phi$ reduces $j_{1x}(\tau,\phi)$ for both $\tau=\pm 1$, similar to the observed behaviors in Figs.\,\ref{fig5}($a$) and \ref{fig5}($b$).
For $j_{1y}(\tau,\phi)$ in Figs.\,\ref{fig9}($c$) and \ref{fig9}($d$), on the other hand, the same Lorentz force initially
strengthens $j_{1y}(\tau,\phi)$ dramatically for all values of $\phi$ and $\tau=\pm 1$ at very low $B_z$
but eventually weakens $j_{1y}(\tau,\phi)$ slowly (quickly) for $\tau=1$ ($\tau=-1$)
in the strong-field limit (in the scale of $\sim 1/B_z$) due to cyclotron motion of electrons.
Such a huge initial increase in $j_{1y}(\tau,\phi)$ at very low $B_z$ is greatly suppressed in graphene with the maximum Berry force at $\phi=0$ (black).
Consequently, a Berry-phase dependent asymmetry in suppressing the skew currents by electron cyclotron motion can be seen by directly comparing Figs.\,\ref{fig9}($c$) with Fig.\,\ref{fig9}($d$).
For a gVHE, the Berry phase can be used for mediating the VHE. In our case, an external magnetic field can be employed further to control this gVHE.
\medskip

\begin{figure}
\centering
\includegraphics[width=0.85\textwidth]{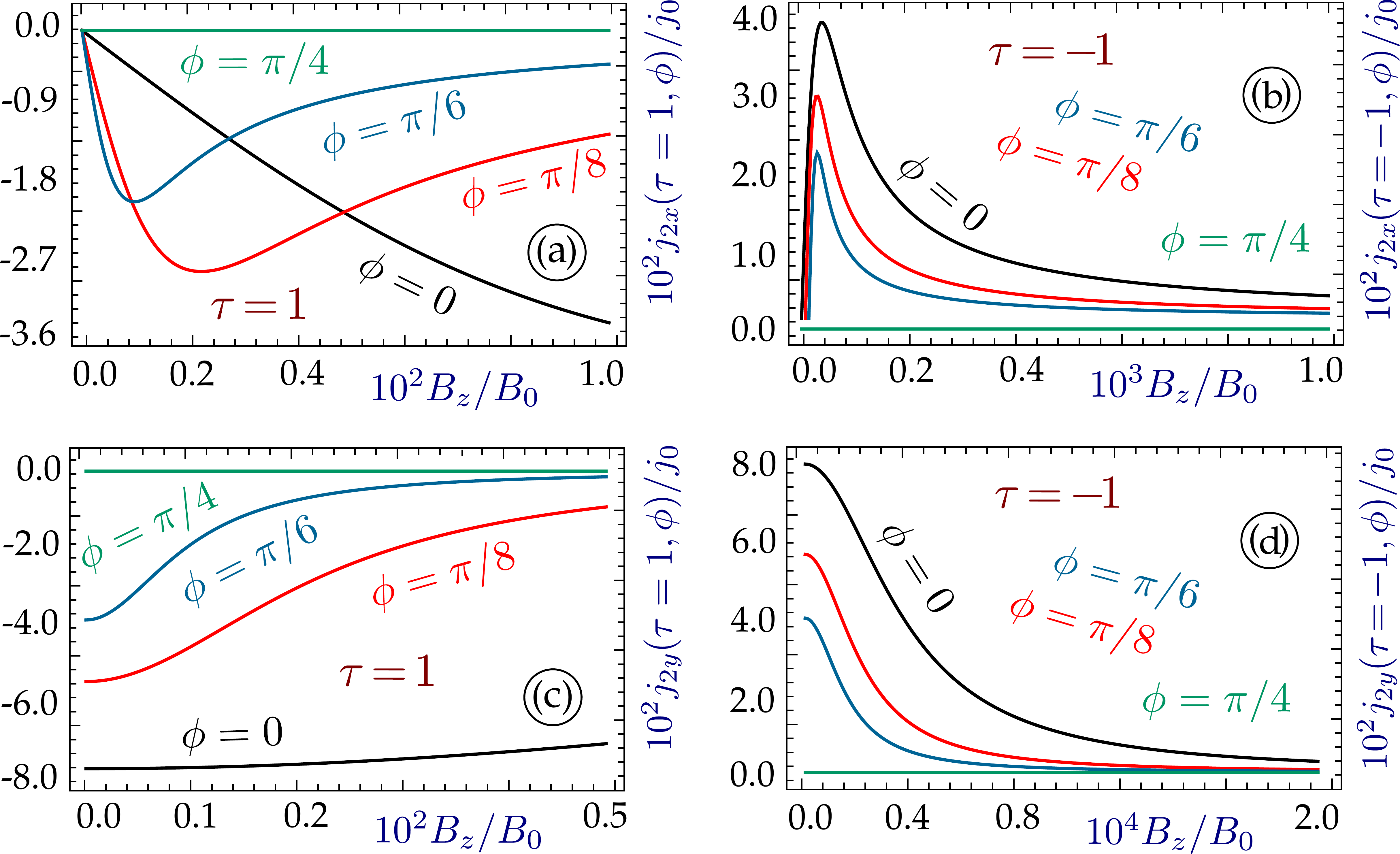}
\caption{Calculated thermal-equilibrium Berry-curvature induced longitudinal current $j_{2x}(\tau,\phi)$ ($a$)-($b$) and Hall current $j_{2y}(\tau,\phi)$ ($c$)-($d$) from Eq.\,(\ref{eqn-6.2})
as a function of $B_z$ with $\phi=\pi/4$ (green), $\phi=\pi/6$ (blue), $\phi=\pi/8$ (red) and $\phi=0$ (black)
for $\tau=1$ ($a$),\,($c$) and $\tau=-1$ ($b$),\,($d$). Here $B_0$ and $j_0$ are given in Figs.\,\ref{fig5} and \ref{fig6}, respectively.}
\label{fig10}
\end{figure}

Finally, from Eq.\,(\ref{eqn-6.2}) we know there exists another conduction current $\mbox{\boldmath$j$}_2(\tau,\phi)$ even in the thermal-equilibrium
state due to Berry curvature $\mbox{\boldmath$\Omega$}_\perp(\mbox{\boldmath$k$}_\|)$, leading to the so-called anomalous Hall effect (AHE) if $\phi\neq\pi/4$.
Figure\ \ref{fig10} presents the calculated AHE current components $j_{2x}(\tau,\phi)$ in ($a$)-($b$) and $j_{2y}(\tau,\phi)$ in ($c$)-($d$).
Since $\mbox{\boldmath$j$}_2(\tau,\phi)$ is proportional to $\tau$ (i.e., valley dependent), we expect the opposite signs in Figs.\,\ref{fig10}($a$) and \ref{fig10}($b$)
for $j_{2x}(\tau,\phi)$ and in Figs.\,\ref{fig10}($c$) and \ref{fig10}($d$) for $j_{2y}(\tau,\phi)$.
As an indication of gVHE,
the increase of the Berry force (or reducing $\phi$) in momentum space will slowly (quickly) enlarge $j_{2x}(\tau,\phi)$ at small $B_z$ and $j_{2y}(\tau,\phi)$ at $B_z=0$ simultaneously due to small (large) mobility at $\tau=1$ ($\tau=-1$).
However, this AHE current is always weakened by the Lorentz force (or increasing $B_z$) in position space for large $B_z$,
where $j_{2x}(\tau,\phi)$ is induced only by one term $\sim B_z\mu_{xx}E_x$, while $j_{2y}(\tau,\phi)$ is generated by two terms
$\sim(1+B_z\mu_{yx})E_x$. Therefore, $j_{2x}(\tau,\phi)$ decreases like $\sim 1/B_z$ in the high-field limit.
Meanwhile, $j_{2y}(\tau,\phi)$ also approaches zero in the same strong-field limit but it scales as $\sim 1/B^2_z$.
Since there are two orders of magnitude difference in $\mu_{xx}$ and $\mu_{yx}$ for $\tau=1$ and $-1$,
we expect the decrease in
$j_{2x}(\tau,\phi)$ and $j_{2y}(\tau,\phi)$ to become much faster at the $\tau=-1$ valley, and therefore
a net AHE current (sum of currents from both valleys) exists and will be dominated by the $\tau=1$ valley for large $B_z$.

\section{Conclusions and Remarks}
\label{sec-4}

In conclusions, we have demonstrated the Berry-phase mediation to valley-dependent Hall transport in $\alpha$-$T_3$ lattices.
We analyze and explain the found interplay between the Lorentz force in position space and the Berry force in momentum space
for the total sheet current density including both normal conduction and Hall currents as well as anomalous Hall current.
We also include many-body screening effects on electron-impurity interactions, which is crucial for avoiding overestimation of elastic scattering.
We further find triplet peak at two distinct valleys and in near-horizontal and near-vertical scattering directions for forward- and back-scattering current,
which favor small Berry phases and low magnetic fields.
We also show a magnetic-field dependence of both non-equilibrium and thermal-equilibrium conduction currents from Berry-phase-mediated and valley-dependent longitudinal and transverse transport.
\medskip

In our theory, we have employed the first two Boltzmann moment equations in calculations of scattering-angle distributions for extrinsic skew-scattering currents due to the presence of random impurities
in $\alpha$-$T_3$ lattices, where both energy- and momentum-relaxation times are computed microscopically.
We attribute this scattering-angle dependence to an anisotropic inverse momentum-relaxation-time tensor calculated within the screened second-order Born approximation and using a static dielectric function within the random-phase approximation.
Meanwhile, we also include the isotropic intrinsic current due to Berry curvature for electrons in thermal-equilibrium states.
Under a perpendicular non-quantizing magnetic field, we find an interplay by Lorentz and valley-dependent resistive forces acting on electrons, leading to field-dependent skew currents.
We further find these skew currents can be mediated by Berry phases of $\alpha$-$T_3$ lattices and depend on barrier- or trap-type
impurity potentials at two inequivalent valleys.

\begin{acknowledgements}
DH would like to acknowledge the financial supports from the Laboratory University Collaboration Initiative (LUCI) program and from the Air Force Office of Scientific Research (AFOSR). Meanwhile, YCL acknowledges financial support from the Vannevar
Bush Faculty Fellowship (VBF) program sponsored by the Basic
Research Office of the Assistant Secretary of Defense for
Research and Engineering and funded by the Office of Naval
Research through Grant No. N00014-16-1-2828.
\end{acknowledgements}

\clearpage
\appendix

\section{Single-Particle Quantum Mechanics}
\label{app-4}

The single-particle Hamiltonian\,\cite{ycl} for an $\alpha$-$T_3$ lattice takes the form of $\tensor{\cal H}_0(\mbox{\boldmath$k$}_\|)=\hbar v_F\tensor{$\mbox{\boldmath${\alpha}$}$}\cdot\mbox{\boldmath$k$}_\|$,
where $\mbox{\boldmath$k$}_\|=\{k_x,k_y\}$, $\tensor{$\mbox{\boldmath${\alpha}$}$}=\{\tensor{\tau_3}\otimes\tensor{S}^\alpha_x,\,\tensor{\tau_0}\otimes\tensor{S}^\alpha_y\}$,
$\tensor{\tau}_{1,2,3}$ are three Pauli matrices, $\tensor{\tau}_0=\tensor{I}_{2\times 2}$ is the identity matrix corresponding to valley degree of freedom,

\begin{equation}
\tensor{S}^\alpha_x=\left[\begin{array}{ccc}
0 & \cos\phi & 0\\
\cos\phi & 0 & \sin\phi\\
0 & \sin\phi & 0
\end{array}\right]\ ,
\ \ \ \ \ \ \ \
\tensor{S}^\alpha_y=\left[\begin{array}{ccc}
0 & -i\cos\phi & 0\\
i\cos\phi & 0 & -i\sin\phi\\
0 & i\sin\phi & 0
\end{array}\right]\ ,
\label{a-3.1}
\end{equation}
and $\alpha=\tan\phi$ ($0\leq\alpha\leq 1$) to parameterize the $\alpha$-$T_3$ lattice.
For this Hamiltonian, three eigenvalues are $\varepsilon_s(k_\|)=s\hbar v_Fk_\|$ with $s=0,\,\pm 1$ as the band index, and the associated eigenstates are

\begin{equation}
|s,\tau,\mbox{\boldmath$k$}_\|\rangle_\phi=\frac{1}{\sqrt{2}}
\left[\begin{array}{c}
\tau\,\cos\phi\,e^{-i\tau\theta_{{\bf k}_\|}}\\
s\\
\tau\,\sin\phi\,e^{i\tau\theta_{{\bf k}_\|}}
\end{array}\right]|\tau\rangle
\label{a-3.2}
\end{equation}
for valley-degenerate eigenvalues $\varepsilon_\pm(k_\|)=\pm\,\hbar v_Fk_\|$ (recorded as ($c$) for $s=+1$ and ($v$) for $s=-1$), and

\begin{equation}
|0,\tau,\mbox{\boldmath$k$}_\|\rangle_\phi=
\left[\begin{array}{c}
\tau\,\sin\phi\,e^{-i\tau\theta_{{\bf k}_\|}}\\
0\\
-\tau\,\cos\phi\,e^{i\tau\theta_{{\bf k}_\|}}
\end{array}\right]|\tau\rangle
\label{a-3.3}
\end{equation}
for $\varepsilon_0(k_\|)=0$, where $\theta_{{\bf k}_\|}=\tan^{-1}(k_y/k_x)$, and $|\tau=\pm 1\rangle$ represent two different valley states.
The Berry connection\,\cite{niu-book} (field) of each band is defined as the quantum-mechanical average of the position operator $\hat{\mbox{\boldmath$r$}}_\|=i\hat{\mbox{\boldmath$\nabla$}}_{{\bf k}_\|}$, i.e., $\mbox{\boldmath$A$}^{\tau,\phi}_s(\mbox{\boldmath$k$}_\|)=\left._\phi\langle s,\tau,\mbox{\boldmath$k$}_\|\right.\vert i\hat{\mbox{\boldmath$\nabla$}}_{{\bf k}_\|}\vert s,\tau,\mbox{\boldmath$k$}_\|\rangle_\phi$ and we get
from Eqs.\,(\ref{a-3.2}) and (\ref{a-3.3})

\begin{equation}
\mbox{\boldmath$A$}^{\tau,\phi}_0(\mbox{\boldmath$k$}_\|)=-\tau\,\frac{1-\alpha^2}{1+\alpha^2}\,\mbox{\boldmath$\nabla$}_{{\bf k}_\|}\theta_{{\bf k}_\|}\ ,\ \ \ \ \ \ \mbox{\boldmath$A$}^{\tau,\phi}_s(\mbox{\boldmath$k$}_\|)=-\frac{1}{2}\,\mbox{\boldmath$A$}^{\tau,\phi}_0(\mbox{\boldmath$k$}_\|)\ .
\end{equation}
Therefore, the Berry curvature $\mbox{\boldmath$\Omega$}^{\tau,\phi}_s(\mbox{\boldmath$k$}_\|)=\mbox{\boldmath$\nabla$}_{{\bf k}_\|}\times\mbox{\boldmath$A$}^{\tau,\phi}_s(\mbox{\boldmath$k$}_\|)$ is calculated as

\begin{equation}
\mbox{\boldmath$\Omega$}^{\tau,\phi}_s(\mbox{\boldmath$k$}_\|)=\tau\left(\frac{1-\alpha^2}{1+\alpha^2}\right)\pi\,\delta(\mbox{\boldmath$k$}_\|)\,\hat{\mbox{\boldmath$e$}}_z\ ,\ \ \ \ \ \ \mbox{\boldmath$\Omega$}^{\tau,\phi}_0(\mbox{\boldmath$k$}_\|)=-2\,\mbox{\boldmath$\Omega$}^{\tau,\phi}_s(\mbox{\boldmath$k$}_\|)\ ,
\end{equation}
where $\hat{\mbox{\boldmath$e$}}_z$ is the unit coordinate vector in the $z$ direction (perpendicular to $\alpha$-$T_3$ plane).

\section{Impurity Scattering Matrix}
\label{app-6}

For impurity scattering of electrons in an $\alpha$-$T_3$ lattice, the initial $\vert i\rangle$ and final $\vert f\rangle$ states for Bloch electrons
with wave vectors $\mbox{\boldmath$k$}_\|$ and $\mbox{\boldmath$k$}'_\|$ can be written as
$\vert i\rangle=\displaystyle{\frac{e^{i{\bf k}_\|\cdot{\bf r}_\|}}{\sqrt{{\cal S}}}}\,\vert s,\tau,\mbox{\boldmath$k$}_\|\rangle_\phi$ and
$\vert f\rangle=\displaystyle{\frac{e^{i{\bf k}'_\|\cdot{\bf r}_\|}}{\sqrt{{\cal S}}}}\,\vert s,\tau,\mbox{\boldmath$k$}'_\|\rangle_\phi$,
where $\vert s,\tau,\mbox{\boldmath$k$}_\|\rangle_\phi$ is given by Eq.\,(\ref{a-3.2}) and ${\cal S}$ is the sheet area.
We assume an isotropic sublattice-selected step-like impurity-scattering potential, i.e., $u^\tau_0(r_\|)=\tau V_0\,\Theta(r_0-r_\|)$, for electrons, where $V_0$ is the step height, $r_0$ represents the interaction range,
and $\tau=+1$ (or $\tau=-1$) corresponds to a barrier-like (or trap-like) impurity potential.
As a result, the screened impurity scattering matrix is found to be\,\cite{defect}

\[
U_{\rm im}^{\tau,\phi}(\mbox{\boldmath$k$}'_\|,\mbox{\boldmath$k$}_\|)=\sum_{{\bf q}'_\|}\,\frac{U^\tau_0(q'_\|)}{\epsilon_\phi(q'_\|)}\,\langle f\vert e^{i{\bf q}'_\|\cdot{\bf r}_\|}\vert i\rangle
=\sum_{{\bf q}'_\|}\,\frac{U^\tau_0(q'_\|)}{\epsilon_\phi(q'_\|)}\,
\]
\[
\times\sum_{\ell}\,\langle f\vert \ell\rangle_{\tau,\phi}\,\left._{\tau,\phi}\langle \ell\right.\vert e^{i{\bf q}'_\|\cdot{\bf r}_\|}\vert i\rangle=\frac{1}{2{\cal S}}\,\sum_{{\bf q}'_\|}\,\frac{U^\tau_0(q'_\|)}{\epsilon_\phi(q'_\|)}\,\sum_{\ell}\,\int\limits_{r'_\|\leq r_0} d^2\mbox{\boldmath$r$}'_\|\,e^{-i{\bf k}'_\|\cdot{\bf r}'_\|}\,\frac{e^{i\ell\Theta_{{\bf r}'_\|}}}{\sqrt{2\pi}}\,
\]
\[
\times\left\{\tau\cos\phi\,e^{-i\tau(\Theta_{{\bf r}'_\|}-\theta_{{\bf k}'_\|})}{\cal R}_1(r'_\|)+s{\cal R}_2(r'_\|)+\tau\sin\phi\,e^{i\tau(\Theta_{{\bf r}'_\|}-\theta_{{\bf k}'_\|})}{\cal R}_3(r'_\|)\right\}\,
\int\limits_{r_\|\leq r_0} d^2\mbox{\boldmath$r$}_\|\,e^{i({\bf q}'_\|+{\bf k}_\|)\cdot{\bf r}_\|}\,
\]
\begin{equation}
\times\frac{e^{-i\ell\Theta_{{\bf r}_\|}}}{\sqrt{2\pi}}\,\left\{\tau\cos\phi\,e^{i\tau(\Theta_{{\bf r}_\|}-\theta_{{\bf k}_\|})}{\cal R}^*_1(r_\|)+s{\cal R}^*_2(r_\|)
+\tau\sin\phi\,e^{-i\tau(\Theta_{{\bf r}_\|}-\theta_{{\bf k}_\|})}{\cal R}^*_3(r_\|)\right\}\ ,
\label{f-1}
\end{equation}
where $U_0(q'_\|)/\epsilon_\phi(q'_\|)$ is the 2D Fourier transform of the screened impurity potential, and

\[
\vert \ell\rangle_{\tau,\phi}=\frac{e^{i\ell\Theta_{{\bf r}_\|}}}{\sqrt{2\pi}}\,\left[
\begin{array}{c}
{\cal R}_1(r_\|)\,e^{-i\tau\Theta_{{\bf r}_\|}}\,\\
{\cal R}_2(r_\|)\\
{\cal R}_3(r_\|)\,e^{i\tau\Theta_{{\bf r}_\|}}\,
\end{array}\right]
\]
are the intermediate quantum states for scattered electrons by an ionized impurity atom with a locally-spherical symmetry [see Eq.\,(\ref{a-7}) below] at the valley $|\tau\rangle$.
Moreover,  the first integral with respect to $\mbox{\boldmath$r$}'_\|$ in Eq.\,(\ref{f-1})
can be evaluated analytically and gives rise to

\[
\mbox{Integral-}\mbox{\boldmath$r$}'_\|=\int\limits_0^{r_0} dr'_\|\,r'_\|\int\limits_0^{2\pi} d\Theta_{{\bf r}'_\|}\,\frac{e^{i\ell\Theta_{{\bf r}'_\|}}}{\sqrt{2\pi}}\,
\sum_{m}\,J_m({k'_\|r'_\|})\,e^{-im(\theta_{{\bf k}'_\|}-\Theta_{{\bf r}'_\|})}\,(-i)^m
\]
\[
\times\left\{\tau\cos\phi\,e^{-i\tau(\Theta_{{\bf r}'_\|}-\theta_{{\bf k}'_\|})}{\cal R}_1(r'_\|)+s{\cal R}_2(r'_\|)+\tau\sin\phi\,e^{i\tau(\Theta_{{\bf r}'_\|}-\theta_{{\bf k}'_\|})}{\cal R}_3(r'_\|)\right\}
\]
\[
=\sqrt{2\pi}(-i)^{\ell}\,e^{i\ell\theta_{{\bf k}'_\|}}\int\limits_0^{r_0} dr'_\|\,r'_\|\left[(-i)^{-\tau}\,\tau\cos\phi\,J_{\ell-\tau}({k'_\|r'_\|})\,{\cal R}_1(r'_\|)\right.
\]
\[
\left.+sJ_{\ell}({k'_\|r'_\|})\,{\cal R}_2(r'_\|)+(-i)^{\tau}\,\tau\sin\phi\,J_{\ell+\tau}({k'_\|r'_\|})\,{\cal R}_3(r'_\|)\right]\ .
\]
Similarly, for the second integral with respect to $\mbox{\boldmath$r$}_\|$ in Eq.\,(\ref{f-1}), we have

\[
\mbox{Integral-}\mbox{\boldmath$r$}_\|=\int\limits_0^{r_0} dr_\|\,r_\|\int\limits_0^{2\pi} d\Theta_{{\bf r}_\|}\,\frac{e^{-i\ell\Theta_{{\bf r}_\|}}}{\sqrt{2\pi}}\,
\sum_{m}\,J_m({|\mbox{\boldmath$k$}_\|+\mbox{\boldmath$q$}'_\||r_\|})\,e^{im(\theta_{{\bf k}_\|+{\bf q}'_\|}-\Theta_{{\bf r}_\|})}\,(i)^m
\]
\[
\times\left\{\tau\cos\phi\,e^{i\tau(\Theta_{{\bf r}_\|}-\theta_{{\bf k}_\|})}{\cal R}^*_1(r_\|)+s{\cal R}^*_2(r_\|)
+\tau\sin\phi\,e^{-i\tau(\Theta_{{\bf r}_\|}-\theta_{{\bf k}_\|})}{\cal R}^*_3(r_\|)\right\}
\]
\[
=\sqrt{2\pi}\,(i)^\ell\,e^{-i\ell\theta_{{\bf k}_\|+{\bf q}'_\|}}\,\int\limits_0^{r_0} dr_\|\,r_\|
\left\{(-i)^\tau\,\tau\cos\phi\,J_{\ell-\tau}(|\mbox{\boldmath$k$}_\|+\mbox{\boldmath$q$}'_\||r_\|)\,{\cal R}^*_1(r_\|)\,e^{i\tau\beta^s_{{\bf k}_\|,{\bf q}'_\|}}\right.
\]
\[
\left.+sJ_\ell(|\mbox{\boldmath$k$}_\|+\mbox{\boldmath$q$}'_\||r_\|)\,{\cal R}^*_2(r_\|)
+(-i)^{-\tau}\,\tau\sin\phi\,J_{\ell+\tau}(|\mbox{\boldmath$k$}_\|+\mbox{\boldmath$q$}'_\||r_\|)\,{\cal R}^*_3(r_\|)\,e^{-i\tau\beta^s_{{\bf k}_\|,{\bf q}'_\|}}\right\}\ ,
\]
where $\beta^s_{{\bf k}_\|,{\bf q}'_\|}=\theta_{{\bf k}_\|+{\bf q}'_\|}-\theta_{{\bf k}_\|}$ is the scattering angle.
Finally, by combining the results for these two integrals and inserting them into Eq.\,(\ref{f-1}) we obtain a simple expression

\begin{equation}
U_{\rm im}^{\tau,\phi}(\mbox{\boldmath$k$}_\|+\mbox{\boldmath$q$}_\|,\mbox{\boldmath$k$}_\|)=\frac{U^\tau_0(q_\|)}{\epsilon_\phi(q_\|)\,{\cal S}}\,{\cal F}_{\tau,\phi}(\mbox{\boldmath$k$}_\|,\mbox{\boldmath$q$}_\|)\ ,
\label{f-2}
\end{equation}
where the form factor ${\cal F}_{\tau,\phi}(\mbox{\boldmath$k$}_\|,\mbox{\boldmath$q$}_\|)$ is defined as

\[
{\cal F}_{\tau,\phi}(\mbox{\boldmath$k$}_\|,\mbox{\boldmath$q$}_\|)=\frac{1}{2}\,\sum_{\ell}\left\{(-i)^{-\tau}\tau\cos\phi\,\chi_1(|\mbox{\boldmath$k$}_\|+\mbox{\boldmath$q$}_\||)+s\chi_2(|\mbox{\boldmath$k$}_\|+\mbox{\boldmath$q$}_\||)\right.
\]
\[
\left.+(-i)^\tau\tau\sin\phi\,\chi_3(|\mbox{\boldmath$k$}_\|+\mbox{\boldmath$q$}_\||)\right\}
\left\{(-i)^\tau\tau\cos\phi\,\chi^*_1(|\mbox{\boldmath$k$}_\|+\mbox{\boldmath$q$}_\||)\,
e^{i\tau\beta^s_{{\bf k}_\|,{\bf q}_\|}}+s\chi^*_2(|\mbox{\boldmath$k$}_\|+\mbox{\boldmath$q$}_\||)\right.
\]
\begin{equation}
\left.+(-i)^{-\tau}\tau\sin\phi\,\chi^*_3(|\mbox{\boldmath$k$}_\|+\mbox{\boldmath$q$}_\||)\,
e^{-i\tau\beta^s_{{\bf k}_\|,{\bf q}_\|}}\right\}\ .
\label{f-3}
\end{equation}
Furthermore, we have introduced the notations in Eq.\,(\ref{f-3}), given by

\begin{equation}
\left\{\begin{array}{l}
\chi_1(|\mbox{\boldmath$k$}_\|+\mbox{\boldmath$q$}_\||)\\
\chi_2(|\mbox{\boldmath$k$}_\|+\mbox{\boldmath$q$}_\||)\\
\chi_3(|\mbox{\boldmath$k$}_\|+\mbox{\boldmath$q$}_\||)
\end{array}\right\}
=\sqrt{2\pi}\int\limits_0^{r_0} dr_\|\,r_\|
\left\{\begin{array}{l}
J_{\ell-\tau}(|\mbox{\boldmath$k$}_\|+\mbox{\boldmath$q$}_\||r_\|)\,{\cal R}_1(r_\|)\\
J_{\ell}(|\mbox{\boldmath$k$}_\|+\mbox{\boldmath$q$}_\||r_\|)\,{\cal R}_2(r_\|)\\
J_{\ell+\tau}(|\mbox{\boldmath$k$}_\|+\mbox{\boldmath$q$}_\||r_\|)\,{\cal R}_3(r_\|)
\end{array}\right\}\ ,
\label{f-4}
\end{equation}
where a wave-function normalization factor should be included as shown in Eq.\,(\ref{a-6}).

\section{Dielectric Function}
\label{app-5}

Under the random-phase approximation\,\cite{book}, the dielectric function $\epsilon_\phi(q_\|,\omega)$ for $\alpha$-$T_3$ lattices is calculated as

\begin{equation}
\epsilon_\phi(q_\|,\omega)=1+\left(\frac{e^2}{2\epsilon_0\epsilon_{\rm r}q_\|}\right)\,{\cal Q}_\phi(q_\|,\omega)\ ,
\label{e-1}
\end{equation}
where the polarization function ${\cal Q}_\phi(q_\|,\omega)$ is given by

\begin{equation}
{\cal Q}_\phi(q_\|,\omega)=\frac{2}{{\cal S}}\sum_{\tau,{\bf k}_\|,s,s'}\,{\cal G}^{\tau,\phi}_{s,s'}(\mbox{\boldmath$k$}_\|,\mbox{\boldmath$q$}_\|)\,
\left\{\frac{f_T^{(0)}[\varepsilon_{s'}(|\mbox{\boldmath$k$}_\|+\mbox{\boldmath$q$}_\||)]-f_T^{(0)}[\varepsilon_{s}(k_\|)]}{\hbar(\omega+i0^+)-\varepsilon_{s'}(|\mbox{\boldmath$k$}_\|+\mbox{\boldmath$q$}_\||)
+\varepsilon_{s}(k_\|)}\right\}\ .
\label{e-2}
\end{equation}
Here, the prefactor $2$ comes from the spin degeneracy, ${\cal S}$ is the sheet area, $\varepsilon_{s}(k_\|)=s\hbar v_F k_\|$ for $s=0,\,\pm 1$, $\omega$ is the angular frequency of a probe field,
$f_T^{(0)}(x)=\{1+\exp[(x-u_0)/k_BT]\}^{-1}$ is the Fermi function for electrons in thermal-equilibrium states, $u_0(T)$ is the chemical
potential for doped electrons, and $T$ is
the temperature. In addition, the overlap integral ${\cal G}^{\tau,\phi}_{s,s'}(\mbox{\boldmath$k$}_\|,\mbox{\boldmath$q$}_\|)$ introduced in Eq.\,(\ref{e-2}) is defined by

\begin{equation}
{\cal G}^{\tau,\phi}_{s,s'}(\mbox{\boldmath$k$}_\|,\mbox{\boldmath$q$}_\|)={\cal G}^{\tau,\phi}_{s',s}(\mbox{\boldmath$q$}_\|,\mbox{\boldmath$k$}_\|)=\left|\left._\phi\Big\langle s,\tau,\mbox{\boldmath$k$}_\|\Big| s',\tau,\mbox{\boldmath$k$}_\|+\mbox{\boldmath$q$}_\|\Big\rangle_\phi\right.\right|^2\ ,
\label{e-3}
\end{equation}
and the wave functions $\vert s,\tau,\mbox{\boldmath$k$}_\|\rangle_\phi$ for $s=0,\,\pm 1$ and $\tau=\pm 1$ are given by Eqs.\,(\ref{a-3.2}) and (\ref{a-3.3}).
At low $T$, the remaining nonzero terms in Eq.\,(\ref{e-2}) in the summation over $s$ and $s'$ correspond to $s'=+1$, $s=0,\,\pm 1$, or vice versa. Therefore, we get three finite terms\,\cite{dice} from Eq.\,(\ref{e-3}):

\begin{equation}
{\cal G}^{\tau,\phi}_{0,+1}(\mbox{\boldmath$k$}_\|,\mbox{\boldmath$q$}_\|)=\frac{1}{2}\,\sin^2(2\phi)\,\sin^2(\beta^s_{{\bf k}_\|,{\bf q}_\|})\ ,
\label{e-4}
\end{equation}

\begin{equation}
{\cal G}^{\tau,\phi}_{\pm 1,+1}(\mbox{\boldmath$k$}_\|,\mbox{\boldmath$q$}_\|)=\frac{1}{4}\,\left\{1\pm\cos(\beta^s_{{\bf k}_\|,{\bf q}_\|})\right\}^2+\frac{1}{4}\,\cos^2(2\phi)\,\sin^2(\beta^s_{{\bf k}_\|,{\bf q}_\|})\ ,
\label{e-5}
\end{equation}
which are independent of $\tau=\pm 1$, where $\beta^s_{{\bf k}_\|,{\bf q}_\|}=\theta_{{\bf k}_\|+{\bf q}_\|}-\theta_{{\bf k}_\|}$ is the angle between two wave vectors $\mbox{\boldmath$k$}_\|$ and $\mbox{\boldmath$k$}_\|+\mbox{\boldmath$q$}_\|$,
and $\theta_{{\bf k}_\|}=\tan^{-1}(k_y/k_x)$ is the angle between $\mbox{\boldmath$k$}_\|$ and $x$-axis.
\medskip

After setting $\omega=0$, we obtain the static dielectric function $\epsilon_\phi(q_\|)$ from Eq.\,(\ref{e-1}) using

\begin{equation}
{\cal Q}_\phi(q_\|,\omega=0)=a_\phi(q_\|)+\Theta(q_\|-2k_F)\,b_\phi(q_\|)\ ,
\label{e-6}
\end{equation}
where $k_F=\sqrt{\pi\rho_0}$, $\rho_0$ is the areal density of doped electrons. If $q_\|<2k_F$ is further assumed, we find $q_\phi=(e^2/2\epsilon_0\epsilon_r)\,a_\phi(q_\|)\approx (e^2k_F)/(\pi\epsilon_0\epsilon_r\hbar v_F)$
for $\epsilon_\phi(q_\|)=1+q_\phi/q_\|$.
As $q_\|\ll k_F$, $a_\phi(q_\|)$ becomes independent of $\phi$ and is given by\,\cite{dice}

\begin{equation}
a_\phi(q_\|)=\frac{1}{2\pi\hbar v_F}\,\left(4k_F+\frac{q_\|^2}{k_F}\right)\approx \frac{2k_F}{\pi\hbar v_F}\ .
\label{e-7}
\end{equation}

\section{Energy-Relaxation Time}
\label{app-1}

By using the detailed-balance condition, the microscopic energy-relaxation time $\tau_\phi(\mbox{\boldmath$k$}_\|,\tau)$ introduced in Eq.\,(\ref{eqn-2}) can be calculated according to\,\cite{huang}

\begin{equation}
\frac{1}{\tau_\phi(\mbox{\boldmath$k$}_\|,\tau)}={\cal W}^{\tau,\phi}_{\rm in}(\mbox{\boldmath$k$}_\|)+{\cal W}^{\tau,\phi}_{\rm out}(\mbox{\boldmath$k$}_\|)\ ,
\label{a-1}
\end{equation}
where the scattering-in rate for electrons in the final $\mbox{\boldmath$k$}_\|$-state is

\begin{equation}
{\cal W}^{\tau,\phi}_{\rm in}(\mbox{\boldmath$k$}_\|)=\frac{\pi N_i}{\hbar}\,\sum_{{\bf q}_\|}\,\left|U^{\tau,\phi}_{\rm im}(\mbox{\boldmath$q$}_\|,\mbox{\boldmath$k$}_\|)\right|^2\,
\left\{f_{{\bf k}_\|-{\bf q}_\|}\,\delta(\varepsilon_{{\bf k}_\|}-\varepsilon_{{\bf k}_\|-{\bf q}_\|})+f_{{\bf k}_\|+{\bf q}_\|}\,
\delta(\varepsilon_{{\bf k}_\|}-\varepsilon_{{\bf k}_\|+{\bf q}_\|})\right\}\ ,\ \
\label{a-2}
\end{equation}
and the scattering-out rate for electrons in the initial $\mbox{\boldmath$k$}_\|$-state is

\[
{\cal W}^{\tau,\phi}_{\rm out}(\mbox{\boldmath$k$}_\|)=\frac{\pi N_i}{\hbar}\,\sum_{{\bf q}_\|}\,\left|U^{\tau,\phi}_{\rm im}(\mbox{\boldmath$q$}_\|,\mbox{\boldmath$k$}_\|)\right|^2\,
\left\{(1-f_{{\bf k}_\|+{\bf q}_\|})\,\delta(\varepsilon_{{\bf k}_\|+{\bf q}_\|}-\varepsilon_{{\bf k}_\|})\right.
\]
\begin{equation}
\left.+(1-f_{{\bf k}_\|-{\bf q}_\|})\,
\delta(\varepsilon_{{\bf k}_\|-{\bf q}_\|}-\varepsilon_{{\bf k}_\|})\right\}\ .
\label{a-3}
\end{equation}
Here, for simplicity, we have introduced the notations $f_{{\bf k}_\|}\equiv f_T^{(0)}[\varepsilon(k_\|)]$ and
$\varepsilon_{{\bf k}_\|}\equiv\varepsilon_+(k_\|)$. We have also assumed low $T$ and $\rho_0$ so that both phonon and pair scattering can be neglected
in comparison with dominant impurity scattering.
In addition, $N_i$ represents the number of randomly-distributed ionized impurities in the system, and $\left|U^{\tau,\phi}_{\rm im}(\mbox{\boldmath$q$}_\|,\mbox{\boldmath$k$}_\|)\right|^2$ comes from the random-impurity scattering
within the second-order Born approximation.
\medskip

Explicitly, using the results in Appendix\ \ref{app-6}, we write down the expression for the screened impurity scattering interaction as

\begin{equation}
\left|U^{\tau,\phi}_{\rm im}(\mbox{\boldmath$q$}_\|,\,\mbox{\boldmath$k$}_\|)\right|^2=\left|\frac{U^\tau_0(q_\|)}{\epsilon_\phi(q_\|)\,{\cal S}}\right|^2\left|{\cal F}_{\tau,\phi}(\mbox{\boldmath$k$}_\|,\mbox{\boldmath$q$}_\|)\right|^2\ ,
\label{a-4}
\end{equation}
where ${\cal S}$ is the sheet area, and $\epsilon_\phi(q_\|)$ is a static dielectric function [see Eqs.\,(\ref{e-1}) and (\ref{e-6})]. In addition,
the scattering form factor ${\cal F}_{\tau,\phi}({\bf k}_\|,{\bf q}_\|)$ in Eq.\,(\ref{a-4}) is given by

\[
{\cal F}_{\tau,\phi}(\mbox{\boldmath$k$}_\|,\mbox{\boldmath$q$}_\|)=\frac{1}{2}\,\sum_{\ell}\left\{(-i)^{-\tau}\tau\cos\phi\,\chi_1(|\mbox{\boldmath$k$}_\|+\mbox{\boldmath$q$}_\||)+s\chi_2(|\mbox{\boldmath$k$}_\|+\mbox{\boldmath$q$}_\||)\right.
\]
\[
\left.+(-i)^\tau\tau\sin\phi\,\chi_3(|\mbox{\boldmath$k$}_\|+\mbox{\boldmath$q$}_\||)\right\}
\left\{(-i)^\tau\tau\cos\phi\,\chi^*_1(|\mbox{\boldmath$k$}_\|+\mbox{\boldmath$q$}_\||)\,
e^{i\tau\beta^s_{{\bf k}_\|,{\bf q}_\|}}+s\chi^*_2(|\mbox{\boldmath$k$}_\|+\mbox{\boldmath$q$}_\||)\right.
\]
\begin{equation}
\left.+(-i)^{-\tau}\tau\sin\phi\,\chi^*_3(|\mbox{\boldmath$k$}_\|+\mbox{\boldmath$q$}_\||)\,
e^{-i\tau\beta^s_{{\bf k}_\|,{\bf q}_\|}}\right\}\ .
\label{a-5}
\end{equation}
where $s=+1$ is selected for doped electrons, $\tau=\pm 1$ for two inequivalent valleys,
$\alpha=\tan\phi$ is the parameter identifying non-equivalent crystalline sublattices,
$\beta^s_{{\bf k}_\|,{\bf q}_\|}\equiv\theta_{{\bf k}_\|+{\bf q}_\|}-\theta_{{\bf k}_\|}$ is the scattering angle, $\theta_{{\bf k}_\|}=\tan^{-1}(k_y/k_x)$, and $\theta_{{\bf k}_\|+{\bf q}_\|}=\tan^{-1}[(k_y+q_y)/(k_x+q_x)]$.
Furthermore, we define the scattering factors in Eq.\,(\ref{a-5}) by

\[
\frac{1}{\sqrt{2\pi}}
\left\{\begin{array}{l}
\chi_1(|\mbox{\boldmath$k$}_\|+\mbox{\boldmath$q$}_\||)\\
\chi_2(|\mbox{\boldmath$k$}_\|+\mbox{\boldmath$q$}_\||)\\
\chi_3(|\mbox{\boldmath$k$}_\|+\mbox{\boldmath$q$}_\||)
\end{array}\right\}
=\left\{\int\limits_0^{1} d\xi\,\xi\left(\left|{\cal R}_1(\xi)\right|^2+\left|{\cal R}_2(\xi)\right|^2+\left|{\cal R}_3(\xi)\right|^2\right)\right\}^{-1/2}
\]
\begin{equation}
\times\int\limits_0^{1} d\xi\,\xi
\left\{\begin{array}{l}
J_{\ell-\tau}(|\mbox{\boldmath$k$}_\|+\mbox{\boldmath$q$}_\||r_0\xi)\,{\cal R}_1(\xi)\\
J_{\ell}(|\mbox{\boldmath$k$}_\|+\mbox{\boldmath$q$}_\||r_0\xi)\,{\cal R}_2(\xi)\\
J_{\ell+\tau}(|\mbox{\boldmath$k$}_\|+\mbox{\boldmath$q$}_\||r_0\xi)\,{\cal R}_3(\xi)
\end{array}\right\}\ ,
\label{a-6}
\end{equation}
where $J_\ell(x)$ is the Bessel function of the first kind, $\ell$ is the angular-momentum quantum number and $r_0$ is the range of impurity interaction.
In addition, the radial parts of the wave function,
${\cal R}_1(\xi)$, ${\cal R}_2(\xi)$ and ${\cal R}_3(\xi)$, introduced in Eq.\,(\ref{a-6}) satisfy the following matrix-form Dirac equation for massless spin-$1$ particles\,\cite{ycl}

\[
\left[
\begin{array}{ccc}
u^\tau_0(\xi) & -\frac{i\tau\hbar v_F\cos\phi}{r_0}\left(\frac{d}{d\xi}+\frac{\tau\ell}{\xi}\right) & 0\\
-\frac{i\tau\hbar v_F\cos\phi}{r_0}\left(\frac{d}{d\xi}-\frac{\tau(\ell-\tau)}{\xi}\right) & u^\tau_0(\xi) & -\frac{i\tau\hbar v_F\sin\phi}{r_0}\left(\frac{d}{d\xi}+\frac{\tau(\ell+\tau)}{\xi}\right)\\
0 & -\frac{i\tau\hbar v_F\sin\phi}{r_0}\left(\frac{d}{d\xi}-\frac{\tau\ell}{\xi}\right) & u^\tau_0(\xi)
\end{array}\right]
\]
\begin{equation}
\bigotimes\left[\begin{array}{c}
{\cal R}_1(\xi)\\ {\cal R}_2(\xi)\\ {\cal R}_3(\xi)
\end{array}\right]=E_0(k_\|)
\left[\begin{array}{c}
{\cal R}_1(\xi)\\ {\cal R}_2(\xi)\\ {\cal R}_3(\xi)
\end{array}\right]\ ,
\label{a-7}
\end{equation}
where $E_0(k_\|)$ represents the given kinetic energy of incident electrons, $u^\tau_0(\xi)=\tau V_0\,\Theta(1-\xi)$ for a barrier-like ($\tau=+1$) or a trap-like ($\tau=-1$) impurity potential,
$V_0$ is a potential-step height in the region of $0\leq\xi=r/r_0\leq 1$, and

\begin{equation}
U_0^\tau(q_\|)=\tau V_0(2\pi r_0^2)\int\limits_0^{1} d\xi\,\xi J_0(\xi r_0q_\|)\ ,
\end{equation}
is the Fourier transform of the scattering potential $u^\tau_0(\xi)$.
It is clear from Eqs.\,(\ref{a-5})-(\ref{a-7}) that ${\cal F}_{\tau,\phi}(\mbox{\boldmath$k$}_\|,\mbox{\boldmath$q$}_\|)\neq{\cal F}_{-\tau,\phi}(\mbox{\boldmath$k$}_\|,\mbox{\boldmath$q$}_\|)$ and
$\chi_1(|\mbox{\boldmath$k$}_\|+\mbox{\boldmath$q$}_\||)\neq\chi_3(|\mbox{\boldmath$k$}_\|+\mbox{\boldmath$q$}_\||)$ if $\phi\neq\pi/4$, which gives rise to valley-dependent impurity scattering.
This can be attributed to the change from the translational symmetry in a crystal to locally-rotational symmetry around an impurity atom.,
as well as to the valley-dependent barrier- or trap-like impurity potential.
\medskip

The matrix-form Dirac equation in Eq.\,(\ref{a-7}) can be solved analytically\,\cite{ycl}, yielding the solutions for $\xi\leq 1$

\begin{equation}
\left[\begin{array}{c}
{\cal R}^\tau_{1,\ell}(\xi)\\ {\cal R}^\tau_{2,\ell}(\xi)\\ {\cal R}^\tau_{3,\ell}(\xi)
\end{array}\right]=
\left[\begin{array}{c}
\cos\phi\,J_{\ell-\tau}(\xi\eta^\tau_0)\\ iS^\tau_0\,J_{\ell}(\xi\eta^\tau_0)\\ -\sin\phi\,J_{\ell+\tau}(\xi\eta^\tau_0)
\end{array}\right]\ ,
\label{soult-2}
\end{equation}
where $\eta^\tau_0(k_\|)=|E_0(k_\|)-\tau V_0|r_0/\hbar v_F$,
and $S^\tau_0={\rm sgn}(E_0(k_\|)-\tau V_0)$ with $(S^\tau_0)^2=1$.
\medskip

Now, we turn to the calculation of $\bar{\tau}_\phi(k_F,\tau)$. From Eq.\,(\ref{a-2}) we get

\begin{equation}
{\cal W}^{\tau,\phi}_{\rm in}(k_\|)=\frac{n_i}{2\pi\hbar^2v_F}\,
k_\|f_{k_\|}\,\sum_{\pm}\int\limits_{-\pi}^{\pi} d\beta_s\,|\cos\theta|\left|\frac{U_0^\tau(2k_\||\cos\theta|)}{\epsilon_\phi(2k_\||\cos\theta|)}\right|^2\,|{\cal F}_{\tau,\phi}(k_\|,\beta_s)|^2\ ,
\label{a-8}
\end{equation}
where $|\cos\theta|=|\sin(|\beta_s|/2)|$, $n_i=N_i/{\cal S}$ is the areal density of ionized impurities, and the summation $\displaystyle{\sum\limits_\pm}$ corresponds to conditions $\varepsilon_{{\bf k}_\|}=\varepsilon_{{\bf k}_\|\pm{\bf q}_\|}$ for two delta-functions in Eq.\,(\ref{a-2}).
Additionally, from Eq.\,(\ref{a-5}) we find for $s=+1$ that

\[
{\cal F}_{\tau,\phi}(k_\|,\beta_s)=\frac{1}{2}\,\sum_{\ell}\left\{(-i)^{-\tau}\tau\cos\phi\,\chi^\tau_{1,\ell}(k_\|)+\chi^\tau_{2,\ell}(k_\|)
+(-i)^\tau\tau\sin\phi\,\chi^\tau_{3,\ell}(k_\|)\right\}
\]
\[
\times\left\{(-i)^\tau\tau\cos\phi\,\chi^\tau_{1,\ell}(k_\|)\,
e^{i\tau\beta^s_{{\bf k}_\|,{\bf q}_\|}}-\chi^\tau_{2,\ell}(k_\|)
+(-i)^{-\tau}\tau\sin\phi\,\chi^\tau_{3,\ell}(k_\|)\,
e^{-i\tau\beta^s_{{\bf k}_\|,{\bf q}_\|}}\right\}
\]
\[
\equiv\kappa_0(k_\|,\phi,\tau)+\kappa_1(k_\|,\phi,\tau)\,e^{i\tau\beta_s}+\kappa_2(k_\|,\phi,\tau)\,e^{-i\tau\beta_s}+\kappa_3(k_\|,\phi,\tau)(1+e^{i\tau\beta_s})
\]
\begin{equation}
+\kappa_4(k_\|,\phi,\tau)(1+e^{-i\tau\beta_s})+\kappa_5(k_\|,\phi,\tau)\cos(\tau\beta_s)\ ,
\label{a-9}
\end{equation}
where

\[
\left\{\begin{array}{l}
\chi^\tau_{1,\ell}(k_\|)\\
\chi^\tau_{2,\ell}(k_\|)\\
\chi^\tau_{3,\ell}(k_\|)
\end{array}\right\}
=\sqrt{2\pi}\left\{\int\limits_0^{1} d\xi\,\xi\left[\cos^2\phi J^2_{\ell-\tau}(\xi\eta^\tau_0)+J^2_{\ell}(\xi\eta^\tau_0)+\sin^2\phi J^2_{\ell+\tau}(\xi\eta^\tau_0)\right]\right\}^{-1/2}
\]
\begin{equation}
\times
\left\{\begin{array}{l}
\cos\phi\\
iS^\tau_0\\
-\sin\phi
\end{array}\right\}
\int\limits_0^{1} d\xi\,\xi
\left\{\begin{array}{l}
J_{\ell-\tau}(k_\|r_0\xi)J_{\ell-\tau}(\xi\eta^\tau_0)\\
J_{\ell}(k_\|r_0\xi)J_{\ell}(\xi\eta^\tau_0)\\
J_{\ell+\tau}(k_\|r_0\xi)J_{\ell+\tau}(\xi\eta^\tau_0)
\end{array}\right\}\ ,
\label{a-20}
\end{equation}
and six real coefficients $\kappa_i$ for $i=0,\,1,\cdots,\,5$ are given by

\begin{eqnarray}
\nonumber
\kappa_0(k_\|,\phi,\tau)&=&\frac{1}{2}\,\sum\limits_{\ell=-\infty}^{\infty}\,\left|\chi^\tau_{2,\ell}(k_\|)\right|^2\ ,\\
\nonumber
\kappa_1(k_\|,\phi,\tau)&=&\frac{1}{2}\cos^2\phi\,\sum\limits_{\ell=-\infty}^{\infty}\,\left[\chi^\tau_{1,\ell}(k_\|)\right]^2\ ,\\
\nonumber
\kappa_2(k_\|,\phi,\tau)&=&\frac{1}{2}\sin^2\phi\,\sum\limits_{\ell=-\infty}^{\infty}\,\left[\chi^\tau_{3,\ell}(k_\|)\right]^2\ ,\\
\nonumber
\kappa_3(k_\|,\phi,\tau)&=&-\frac{i}{2}\cos\phi\,\sum\limits_{\ell=-\infty}^{\infty}\,\chi^\tau_{1,\ell}(k_\|)\chi^\tau_{2,\ell}(k_\|)\ ,\\
\nonumber
\kappa_4(k_\|,\phi,\tau)&=&+\frac{i}{2}\sin\phi\,\sum\limits_{\ell=-\infty}^{\infty}\,\chi^\tau_{2,\ell}(k_\|)\chi^\tau_{3,\ell}(k_\|)\ ,\\
\label{a-17}
\kappa_5(k_\|,\phi,\tau)&=&-\frac{1}{2}\sin 2\phi\,\sum\limits_{\ell=-\infty}^{\infty}\,\chi^\tau_{1,\ell}(k_\|)\chi^\tau_{3,\ell}(k_\|)\ .
\end{eqnarray}
Then, at low $T$, from the detailed-balance condition and Eq.\,(\ref{a-8}) we finally arrive at

\[
\frac{1}{\bar{\tau}_\phi(k_F,\tau)}=\frac{4}{\rho_0{\cal S}}\sum_{{\bf k}_\|}\,\frac{f_T^{(0)}[\varepsilon(k_\|)]}{\tau_\phi(\mbox{\boldmath$k$}_\|,\tau)}=\frac{4}{\rho_0{\cal S}}\sum_{{\bf k}_\|}\,{\cal W}^{\tau,\phi}_{\rm in}(k_\|)\,\Theta(k_F-k_\|)
\]
\begin{equation}
=\frac{4n_i}{\pi^2\hbar^2v_F\rho_0}
\int\limits_{-\pi}^{\pi} d\beta_s\,|\cos\theta|\int\limits_0^{k_F} dk_\|\,k_\|^2\left|\frac{U^\tau_0(2k_\||\cos\theta|)}{\epsilon_\phi(2k_\||\cos\theta|)}\right|^2
|{\cal F}_{\tau,\phi}(k_\|,\beta_s)|^2\ .
\label{a-10}
\end{equation}

\section{Inverse Momentum-Relaxation-Time Tensor}
\label{app-2}

The inverse momentum-relaxation-time tensor $\tensor{\mbox{\boldmath$\cal T$}}_p^{-1}(\tau,\phi)$ introduced in Eq.\,(\ref{eqn-4}) comes from the statistically-averaged resistive forces $\mbox{\boldmath$f$}_i(\tau,\phi)$
due to scattering of electrons by ionized impurities ($i$) at low temperatures.\,\cite{jmo,backes}
\medskip

For electrons moving with a center-of-mass momentum $\hbar\mbox{\boldmath$K$}^{\tau,\phi}_0$, the resistive force $\mbox{\boldmath$f$}_i(\tau,\phi)$ from impurity scattering is calculated as\,\cite{huang}

\[
\mbox{\boldmath$f$}_i(\tau,\phi)=-N_i\left(\frac{2\pi}{\hbar}\right)\frac{v_F}{k_F}\,\sum_{{\bf k}_\|,{\bf q}_\|}\,\hbar\mbox{\boldmath$q$}_\|\left(\hbar\mbox{\boldmath$q$}_\|\cdot\mbox{\boldmath$K$}^{\tau,\phi}_0\right)\,
\]
\begin{equation}
\times\left|U^{\tau,\phi}_{\rm im}(\mbox{\boldmath$q$}_\|,\,\mbox{\boldmath$k$}_\|)\right|^2\,
\left(-\frac{\partial f_{{\bf k}_\|}}{\partial\varepsilon_{{\bf k}_\|}}\right)\,\delta(\varepsilon_{{\bf k}_\|+{\bf q}_\|}-\varepsilon_{{\bf k}_\|})\ ,
\label{b-1}
\end{equation}
and we have $\tensor{\mbox{\boldmath$\cal T$}}_i^{-1}(\tau,\phi)\cdot\mbox{\boldmath$K$}^{\tau,\phi}_0=-\mbox{\boldmath$f$}_i(\tau,\phi)/N_0\hbar$ by definition. This leads to

\begin{equation}
\tensor{\mbox{\boldmath$\cal T$}}_i^{-1}(\tau,\phi)=\frac{2\pi N_iv_F}{N_0k_F}\,\sum_{{\bf k}_\|,{\bf q}_\|}\,\left|U^{\tau,\phi}_{\rm im}(\mbox{\boldmath$q$}_\|,\,\mbox{\boldmath$k$}_\|)\right|^2\,
\left(-\frac{\partial f_{{\bf k}_\|}}{\partial\varepsilon_{{\bf k}_\|}}\right)\,\delta(\varepsilon_{{\bf k}_\|+{\bf q}_\|}-\varepsilon_{{\bf k}_\|})\,
\left[\mbox{\boldmath$q$}_\|\otimes\mbox{\boldmath$q$}_\|^T\right]\ ,
\label{b-2}
\end{equation}
where $\displaystyle{\left[\mbox{\boldmath$q$}_\|\otimes\mbox{\boldmath$q$}_\|^T\right]\equiv\left[\begin{array}{cc}
q_x^2 & q_xq_y\\
q_yq_x & q_y^2
\end{array}\right]\ .
}$
Finally, the inverse momentum-relaxation-time tensor is simply given by $\tensor{\mbox{\boldmath$\cal T$}}_p^{-1}(\tau,\phi)=\tensor{\mbox{\boldmath$\cal T$}}_i^{-1}(\tau,\phi)$ after neglecting phonon scattering at low $T$.
\medskip

Furthermore, at low $T$, from Eqs.\,(\ref{b-1}) and (\ref{b-2}) we find

\[
\tensor{\mbox{\boldmath$\cal T$}}_p^{-1}(k_F,\tau,\phi)=\frac{2\pi n_i}{\rho_0}\,\left(\frac{v_F}{k_F}\right)\sum_{{\bf k}_\|,{\bf q}_\|}\,\left|U^{\tau,\phi}_{\rm im}(\mbox{\boldmath$q$}_\|,\,\mbox{\boldmath$k$}_\|)\right|^2\,
\delta(\varepsilon_{{\bf k}_\|}-E_F)\,\delta(\varepsilon_{{\bf k}_\|+{\bf q}_\|}-\varepsilon_{{\bf k}_\|})\,
\left[\mbox{\boldmath$q$}_\|\otimes\mbox{\boldmath$q$}_\|^T\right]
\]
\[
=\frac{4n_ik^3_F}{\pi^2\hbar^2v_F\rho_0}
\int\limits_{-\pi}^{\pi} d\beta_s\,|\cos\theta|\cos^2\theta\left|\frac{U^\tau_0(2k_F|\cos\theta|)}{\epsilon^2_\phi(2k_F|\cos\theta|)}\right|^2|{\cal F}_{\tau,\phi}(k_F,\beta_s)|^2
\]
\begin{equation}
\times\left[\begin{array}{cc}
\cos^2\theta & \cos\theta\sin\theta\\
\sin\theta\cos\theta & \sin^2\theta
\end{array}\right]\ ,
\label{b-3}
\end{equation}
where $\epsilon_\phi(q_\|)$ is the static dielectric function, $|{\cal F}_{\tau,\phi}(k_F,\beta_s)|^2$ is given by Eq.\,(\ref{a-9}),
$\cos\theta=-\sin(|\beta_s|/2)$, $\sin\theta={\rm sgn}(\beta_s)\,\cos(|\beta_s|/2)$ for $-\pi\leq\beta_s\leq\pi$, and ${\rm sgn}(x)$ is a sign function.

\section{Mobility Tensor}
\label{app-3}

From the force-balance equation in Eq.\,(\ref{eqn-4}), we get the following set of linear equations\,\cite{backes} for center-of-mass wave vector $\mbox{\boldmath$K$}^{\tau,\phi}_0=\{K^{\tau,\phi}_x,K^{\tau,\phi}_y\}$, i.e.,

\begin{equation}
b_{xx}(\tau,\phi)K^{\tau,\phi}_x+\left[b_{xy}(\tau,\phi)-\frac{q_0v_FB_z}{\hbar k_F}\right]K^{\tau,\phi}_y
=\frac{q_0}{\hbar}\,E_x\ ,
\label{dan-58}
\end{equation}

\begin{equation}
\left[b_{yx}(\tau,\phi)+\frac{q_0v_FB_z}{\hbar k_F}\right]K^{\tau,\phi}_x+b_{yy}(\tau,\phi)K^{\tau,\phi}_y
=\frac{q_0}{\hbar}\,E_y\ ,
\label{dan-59}
\end{equation}
where we have used the notations $\mbox{\boldmath$B$}_\perp=\{0,0,B_z\}$, $\mbox{\boldmath$E$}_\|=\{E_x,E_y,0\}$, $q_0=-e$,
and have written the matrix $\tensor{\mbox{\boldmath$\cal T$}}_p^{-1}(\tau,\phi)\equiv\{b_{ij}(\tau,\phi)\}$ for $i,j=x,y$.
By defining the determinant of the coefficient matrix in Eqs.\,\eqref{dan-58} and \eqref{dan-59}
as $Det\{\tensor{\mbox{\boldmath$\cal C$}}_{\tau,\phi}\}$, i.e.,

\begin{equation}
Det\{\tensor{\mbox{\boldmath$\cal C$}}_{\tau,\phi}\}=
b_{xx}(\tau,\phi)\,b_{yy}(\tau,\phi)-\left[b_{xy}(\tau,\phi)-\frac{q_0v_FB_z}{\hbar k_F}\right]\left[b_{yx}(\tau,\phi)+\frac{q_0v_FB_z}{\hbar k_F}\right]\ ,
\label{dan-61}
\end{equation}
as well as the source vector $\mbox{\boldmath$s$}$, given by

\begin{equation}
\mbox{\boldmath$s$}=\left[\begin{array}{c}
\displaystyle{\frac{q_0}{\hbar}\,E_x}\\
\displaystyle{\frac{q_0}{\hbar}\,E_y}
\end{array}\right]\ ,
\label{dan-62}
\end{equation}
we can reduce this linear equations to a matrix form $\tensor{\mbox{\boldmath$\cal C$}}_{\tau,\phi}\cdot\mbox{\boldmath$K$}^{\tau,\phi}_0=\mbox{\boldmath$s$}$ with the formal solution
$\mbox{\boldmath$K$}^{\tau,\phi}_0=\tensor{\mbox{\boldmath$\cal C$}}_{\tau,\phi}^{-1}\cdot\mbox{\boldmath$s$}$. Explicitly, we find the solution $\mbox{\boldmath$K$}^{\tau,\phi}_0=\{K^{\tau,\phi}_x,K^{\tau,\phi}_y\}$
for $j=x,y$ from

\begin{equation}
K^{\tau,\phi}_j=\frac{Det\{\tensor{\mbox{\boldmath$\Delta$}}^{\tau,\phi}_j\}}{Det\{\tensor{\mbox{\boldmath$\cal C$}}_{\tau,\phi}\}}\ ,
\label{dan-63}
\end{equation}
where

\begin{equation}
Det\{\tensor{\mbox{\boldmath$\Delta$}}^{\tau,\phi}_1\}=\frac{q_0}{\hbar}\,E_x\,b_{yy}(\tau,\phi)
-\frac{q_0}{\hbar}\,E_y\left[b_{xy}(\tau,\phi)-\frac{q_0v_FB_z}{\hbar k_F}\right]\ ,
\label{dan-64}
\end{equation}

\begin{equation}
Det\{\tensor{\mbox{\boldmath$\Delta$}}^{\tau,\phi}_2\}=\frac{q_0}{\hbar}\,E_y\,b_{xx}(\tau,\phi)
-\frac{q_0}{\hbar}\,E_x\left[b_{yx}(\tau,\phi)+\frac{q_0v_FB_z}{\hbar k_F}\right]\ .
\label{dan-65}
\end{equation}
Even in the case of $E_y=0$, the transverse center-of-mass wave number $K^{\tau,\phi}_y$ can still be nonzero due to an external magnetic field $B_z$ or by nonzero off-diagonal element $b_{yx}$ of the inverse momentum-relaxation-time tensor.	
The mobility tensor $\tensor{\mbox{\boldmath$\mu$}}_{\tau,\phi}=\{\mu^{\tau,\phi}_{ij}\}$ can be simply obtained from $\mu^{\tau,\phi}_{ij}=(v_F/k_F)\,(\partial K^{\tau,\phi}_i/\partial E_j)$.

\clearpage

\end{document}